\documentclass[aip,reprint,author-year,nofootinbib,floatfix]{revtex4-2}

\usepackage{physjour_aas_macros}

\usepackage{graphicx,bm}
\usepackage{array}[=2016-10-06] 
\usepackage{amsmath,amssymb}             
\usepackage{stmaryrd}					
\usepackage[utf8]{inputenc}
\usepackage[T1]{fontenc}
\usepackage[dvipsnames]{xcolor}
\usepackage{color}
\usepackage[raggedright]{titlesec}
\usepackage[normalem]{ulem}
\definecolor{darkblue}{rgb}{0.0,0.0,0.4}
\definecolor{darkred}{rgb}{0.7,0.0,0.0}
\definecolor{darkgreen}{rgb}{0.0,0.5,0.0}
\definecolor{C0}{HTML}{1f77b4}
\definecolor{C1}{HTML}{ff7f0e}
\definecolor{C2}{HTML}{2ca02c}
\definecolor{C3}{HTML}{d62728}
\definecolor{C4}{HTML}{9467bd}
\definecolor{C5}{HTML}{8c564b}
\definecolor{C6}{HTML}{e377c2}
\definecolor{C7}{HTML}{7f7f7f}
\definecolor{C8}{HTML}{bcbd22}
\definecolor{C9}{HTML}{17becf}

\usepackage{tikz}
\usetikzlibrary{shapes,arrows,shadows}
\usepackage[export]{adjustbox}

\usepackage{scalerel}
\usetikzlibrary{svg.path}
\definecolor{orcidlogocol}{HTML}{A6CE39}
\tikzset{
  orcidlogo/.pic={
    \fill[orcidlogocol] svg{M256,128c0,70.7-57.3,128-128,128C57.3,256,0,198.7,0,128C0,57.3,57.3,0,128,0C198.7,0,256,57.3,256,128z};
    \fill[white] svg{M86.3,186.2H70.9V79.1h15.4v48.4V186.2z}
                 svg{M108.9,79.1h41.6c39.6,0,57,28.3,57,53.6c0,27.5-21.5,53.6-56.8,53.6h-41.8V79.1z M124.3,172.4h24.5c34.9,0,42.9-26.5,42.9-39.7c0-21.5-13.7-39.7-43.7-39.7h-23.7V172.4z}
                 svg{M88.7,56.8c0,5.5-4.5,10.1-10.1,10.1c-5.6,0-10.1-4.6-10.1-10.1c0-5.6,4.5-10.1,10.1-10.1C84.2,46.7,88.7,51.3,88.7,56.8z};
  }
}
\newcommand\orcid[1]{\href{https://orcid.org/#1}{\mbox{\scalerel*{
\begin{tikzpicture}[yscale=-1,transform shape]
\pic{orcidlogo};
\end{tikzpicture}
}{|}}}}

\usepackage{hyperref}
\hypersetup{colorlinks,breaklinks,
            linkcolor=darkblue,urlcolor=darkblue,
            anchorcolor=darkblue,citecolor=RoyalBlue}

\usepackage[mathscr]{euscript}
\usepackage{upgreek}
\DeclareFontFamily{OT1}{pzc}{}
\DeclareFontShape{OT1}{pzc}{m}{it}{<-> s * [1.10] pzcmi7t}{}
\DeclareMathAlphabet{\mathpzc}{OT1}{pzc}{m}{it}

\titleformat{\section}{\selectfont \normalfont\raggedright\sffamily\small\bfseries\uppercase}{\thesection.}{1em}{}{}
\titleformat{\subsection}{\selectfont \normalfont\raggedright\sffamily\small\bfseries}{\thesubsection.}{1em}{}{}
\titleformat{\subsubsection}{\selectfont \normalfont\sffamily\small\bfseries}{\thesubsection.\thesubsubsection}{1em}{}{}
\titleformat{\paragraph}[runin]{\selectfont \normalfont\sffamily\small\bfseries}{\thesubsection.\thesubsubsection.\theparagraph}{1em}{}[.]

\usepackage{tikz}
\usetikzlibrary{shapes,arrows,shadows}
\usepackage[export]{adjustbox}

\usepackage{multirow}

\usepackage{enumitem}

\newcommand{\deriv}[2]{\dfrac{\mathrm{d} #1}{\mathrm{d} #2}}

\newcommand{\drm}{\mathrm{d}}
\newcommand{\erm}{\mathrm{e}}
\newcommand{\irm}{\mathrm{i}}

\newcommand{\G}{\mathrm{G}}

\begin{document}

\title{Evolving only the observed spacetime region: perfectly parallel lightcone simulations with sCOLA}

\newcommand{\iap}{CNRS \& Sorbonne Université, UMR 7095, Institut d’Astrophysique de Paris, 98 bis boulevard Arago, F-75014 Paris, France}



\author{Mayeul Aubin}
\email{mayeul.aubin@iap.fr}
\thanks{ORCID: \orcid{0009-0004-4673-3864} \href{https://orcid.org/0009-0004-4673-3864}{0009-0004-4673-3864}.}
\affiliation{\iap}

\author{Florent Leclercq}
\email{florent.leclercq@iap.fr}
\homepage{https://www.florent-leclercq.eu/}
\thanks{ORCID: \orcid{0000-0002-9339-1404} \href{https://orcid.org/0000-0002-9339-1404}{0000-0002-9339-1404}.}
\affiliation{\iap}

\author{Guilhem Lavaux}
\email{guilhem.lavaux@iap.fr}
\homepage{https://www.guilhem-lavaux.eu/}
\thanks{ORCID: \orcid{0000-0003-0143-8891} \href{https://orcid.org/0000-0003-0143-8891}{0000-0003-0143-8891}.}
\affiliation{\iap}


\date{\today}

\begin{abstract}
\noindent
Cosmological analyses of Stage-IV galaxy surveys require large ensembles of mock catalogues built from lightcone simulations, but a single full $N$-body run at survey scale costs of order $10^7$ CPU hours, which puts such ensembles out of reach.
We present a framework that produces survey-scale dark matter lightcones for a typical cost of order $10^3$ CPU hours each, based on the spatial COmoving Lagrangian Acceleration (sCOLA) method.
sCOLA decomposes an arbitrarily large volume into small tiles that are evolved independently, with buffer regions and non-periodic boundary conditions on the gravitational potential. We combine it with on-the-fly lightcone building, so that the evolved spacetime region reduces to the observed one: each tile is stopped as soon as the observer's past lightcone has left it, and only the tiles intersecting the region of interest are simulated at all. We further introduce two accuracy improvements to the sCOLA formalism: an LPT buffer that approximates the inflow of particles from outside the sCOLA box, and a second-order perturbative approximation to the gravitational potential at its boundaries; together they remove the large-scale inaccuracies of previous work and allow smaller tiles and buffers.
Against a \texttt{Gadget4} reference in a $3.84\,h^{-1}\,\mathrm{Gpc}$ box with $3\,840^3$ particles, up to $k \simeq 0.4\,h\,\mathrm{Mpc}^{-1}$ the sCOLA lightcone reproduces the matter power spectrum to 1\%, the velocity divergence power spectrum to 1.5\% and the equilateral bispectrum to 2\%, with no visible discontinuity at tile boundaries. The run cost $1\,300$ CPU hours, a factor of $2\,600$ below our \texttt{Gadget4} run.
Perfect parallelism at this cost brings ensembles of thousands of survey-scale lightcones within reach of a single allocation: $\sim 25\,000$ realisations of the \textit{Euclid} Galactic north cap with a mass resolution of $10^{11}\,\mathrm{M}_\odot$ for $10^{7}$ CPU hours.
\end{abstract}

\maketitle


\section{Introduction}

    Stage-IV galaxy surveys such as \textit{Euclid} \citep{EuclidForecast,EuclidForecast2}, DESI \citep{DESIProgram}, the Rubin Observatory Legacy Survey of Space and Time \citep{Ivezic2019}, \textit{Roman} \citep{Spergel2015} and \textit{Xuntian} \citep{Gong2026} map the large-scale structure of the Universe over tens of cubic gigaparsecs and over most of cosmic history. Their cosmological constraints increasingly come from the full shape of the clustering statistics rather than from a few compressed summaries \citep{SDSS_FullShape,DESI_FullShape}, which makes them sensitive to the non-linear regime of structure formation, where no exact analytical description exists.

    Interpreting these data therefore relies on numerical $N$-body simulations of gravitational collapse \citep{Review_Simulations}. The mock catalogues built from them calibrate an analysis pipeline end to end, provide the covariance matrices of the summary statistics, whose noise otherwise biases the inferred parameters \citep{Hartlap2007,DodelsonSchneider2013,Taylor2013,DESI_covariance}, and serve as the forward model of simulation-based \citep{Makinen2022,deSanti2023,SimBig} and field-level \citep{Elsner2020,Doeser2023,Nguyen2024,LULO} inference. What a survey observes, however, is not the Universe at one epoch but the observer's past lightcone: each object is seen at the time when the light we receive from it was emitted. A mock catalogue is accordingly assembled in stages --- evolving the dark matter, identifying halos \citep[e.g.][]{Rockstar}, assigning galaxies to them by a prescription calibrated on observations \citep[e.g.][]{Wechsler2018}, and placing those galaxies on the dark matter lightcone \citep[e.g.][]{Euclid_skies} --- of which only the first is expensive: the others are comparatively cheap, and it is the cost of the dark matter evolution that decides whether an ensemble of mocks can be produced at all. The present paper is concerned with that step alone: producing the dark matter lightcone.

    The scale required for these mock catalogues is set by the surveys themselves. For example, the Galactic north cap covered by the \textit{Euclid} spectroscopic sample spans the redshift range $0.84 \leq z \leq 1.88$ and a comoving volume of tens of $\mathrm{Gpc}^3$ \citep{Euclid_Wide_Survey}. A lightcone of that region must cover the same range of epochs, and resolve the halos that host the observed tracers. One simulation of that size is already a major undertaking: state-of-the-art projects consume allocations of order $10^7$ CPU hours \citep{EuclidMock,Euclid_skies,HalfDome}. Covariance estimation and simulation-based inference, however, require not one such realisation but thousands, and at that unit cost an ensemble is out of reach. What is needed is a method that delivers one survey-scale lightcone for a cost of order $10^3$ CPU hours.

    The lightcones used today are produced with full $N$-body codes such as \texttt{PKDGRAV3} \citep{PKDGRAV}, \texttt{Gadget4} \citep{Gadget4}, \texttt{RAMSES} \citep{RAMSES}, \texttt{GreeM} \citep{GreeM}, \texttt{Abacus} \citep{AbacusCode} and \texttt{HACC} \citep{HACCDataRelease}, which have produced the landmark runs of the field: the Horizon 4$\pi$ simulation \citep{Teyssier2009}, the DEUS Full Universe Run \citep{Alimi2012}, MICE-GC \citep{MICEGC3,Lightcone_onion2,MICEGC2}, the \textit{Euclid Flagship} simulation \citep{EuclidMock}, Uchuu \citep{Uchuu}, AbacusSummit \citep{AbacusSummit} and Farpoint \citep{Farpoint_sim}. They reach the accuracy the surveys demand, but at a unit cost that rules out large ensembles: most of these projects deliver a single realisation of the Universe. Such a realisation can be compared with the data, but it cannot on its own furnish the ensemble statistics an analysis needs (a covariance matrix, or the distribution of a summary statistic over realisations) which require many independent realisations of the same volume.

    Faster alternatives trade accuracy for speed. \texttt{Pinocchio} \citep{Pinocchio_new,masspeak_patch}, used for the Gepetto mocks \citep{Euclid_skies}, predicts halo formation from ellipsoidal collapse instead of solving the equations of motion at all. Where the equations of motion are kept, two distinct strategies reduce the number of force evaluations. The first is to build a time integrator informed by perturbation theory, so that a handful of steps suffices: this is the approach of \texttt{FastPM} \citep{FastPM}, of the integrators of \citet{List2024}, and of \texttt{BullFrog} \citep{Bullfrog}. The second is to change the frame of reference: COLA \citep[COmoving Lagrangian Acceleration,][]{Tassev2013} and COCA \citep[COmoving Computer Acceleration,][]{Bartlett2025tCOCA} solve for the residual displacement with respect to an approximation of the real trajectories of particles.
    In particular, COLA has already been used to produce mocks: \texttt{cola\_halo} \citep{Koda2016} fills the survey volume with remapped and replicated snapshots in redshift bins, while \texttt{L-PICOLA} \citep{LPICOLA} and \texttt{ICE-COLA} \citep{Izard2016,Izard2018} build the lightcone at run time. Machine-learning emulators go further still, replacing the dynamics by a map learnt from the initial conditions \citep{ML_emulator,Jamieson2023,ML_emulator2,Doeser2023}; most of them return snapshots at fixed redshift. Two recent exceptions are a generative model of halo lightcones \citep{Zacharegkas2026} and a Lagrangian neural cellular automaton that integrates particle trajectories in time, from which a matter lightcone can be recorded on the fly \citep{LNCA}. Whichever approximation is adopted, the small-scale clustering of the resulting lightcone mocks departs from a full $N$-body reference and has to be corrected \citep{Lightcone_assessment}.
         
    None of these methods addresses the inefficiency specific to lightcones. A lightcone is usually assembled by stacking replicated copies of the box into concentric shells \citep{Blaizot2005,Kitzbichler2007,Lightcone_onion,Lightcone_onion2}, or by recording lightcone crossings on the fly and interpolating between time steps \citep{Evrard2002,Lightcone_interpolation,Lightcone_interpolation2,LANTERN,Ramakrishnan2025}. Both methods must replicate the box whenever the simulated volume is smaller than the observed one \citep[e.g.][]{LPICOLA,Izard2018}, as is the case even for the largest runs to date: the \textit{Euclid} Flagship simulation replicates its $3.6\,h^{-1}\,\mathrm{Gpc}$ box to cover the sky out to $z=3$ \citep{EuclidMock}, HalfDome tiles its $3.75\,h^{-1}\,\mathrm{Gpc}$ box out to $z=4$ \citep{HalfDome}, and the Geppetto mocks tile a $1.2\,h^{-1}\,\mathrm{Gpc}$ box out to $z=2$ \citep{Euclid_skies}. The same structures then recur in different directions and at different redshifts, biasing higher-order statistics and small-area surveys notably; 
    reorienting the lightcone inside the box only delays the first repetition \citep{BoxReplicationWL,Ramakrishnan2025}. Both approaches also evolve the whole volume to redshift zero, long after distant particles crossed the lightcone: much of the computation is spent on particles already observed. Shrinking the simulation domain as the lightcone contracts recovers part of that waste \citep{Lightcone_shrinkingdomain,Lightcone_shrinkingdomain2}, at the price of reshaping it during the run.

    In this paper we use sCOLA \citep{Tassev2015,Leclercq2020sCOLA}, an extension of COLA that splits an arbitrarily large volume into small tiles evolved independently (here $216\,000$ tiles of side $64\,h^{-1}\,\mathrm{Mpc}$). 
    That independence is what makes the tiling well suited to lightcone construction. A tile can be stopped as soon as the past lightcone has left it, each of its particles having been recorded at its own crossing time, without any reference to the rest of the volume --- a restriction in time. And only the tiles intersecting the survey footprint need be evolved at all --- a restriction in space.
 Together, they make the evolved spacetime region coincide with the observed one, up to the tile buffers: no computation is spent on particles that have already been observed, nor on volume that will never be. What dedicated lightcone schemes obtain by reshaping the simulated domain during the run is therefore obtained here at no cost, as a by-product of the decomposition. Measured against a full $N$-body reference run with \texttt{Gadget4} from the same initial conditions, an sCOLA lightcone is about three orders of magnitude cheaper, at the price of approximate dynamics below the scales resolved by the particle mesh. We also introduce two improvements of the sCOLA formalism of \citet{Leclercq2020sCOLA} at the boundaries of the tiles, where its approximations are least accurate. Together they allow substantially smaller buffers, and therefore a smaller fraction of the computation spent on volume that is afterwards discarded.

    Section~\ref{sec:Methods} recalls the sCOLA paradigm, introduces our improvements to it, and details the lightcone-building procedure. Section~\ref{sec:Simulations} describes the simulations we ran. Section~\ref{sec:Results} reports the accuracy and the cost of the method, which we discuss in Section~\ref{sec:Discussion}. We conclude in Section~\ref{sec:Conclusion}; implementation details and parameter studies are provided in the appendices.

\section{Methods}
\label{sec:Methods}

We begin by reviewing the tCOLA and sCOLA methods in Sections~\ref{ssec:Review of tCOLA} and \ref{ssec:Review of sCOLA}, respectively. In Section~\ref{ssec:Improvements of sCOLA}, we discuss the improvements to sCOLA accuracy introduced by this paper, and in Section~\ref{ssec:Lightcone building} we describe our lightcone-building procedure.

    \subsection{Review of tCOLA}
    \label{ssec:Review of tCOLA}

The temporal COmoving Lagrangian Acceleration \citep[tCOLA,][]{Tassev2013} method is a hybrid approach designed to combine the accuracy of $N$-body simulations with the efficiency of analytical approximations. In this framework, the equations of motion for a particle are solved in a reference frame that co-moves with the approximate trajectory of the particle, as predicted by Lagrangian Perturbation Theory (LPT).

In the following, we adopt the notations of \citet{LeclercqThesis,Leclercq2020sCOLA,Bartlett2025tCOCA}, where complete derivations can be found. Let $\tau$ denote the conformal time and $a$ denote the scale factor, normalised to unity at the present time, i.e., $a_0 = 1$. The initial and final comoving positions of particles are denoted by $\bm{q}$ and $\bm{x}$, respectively. The canonical momenta with respect to the Hubble flow, scaled by $a^2$, are defined as
\begin{equation}
\bm{p} \equiv a \frac{\drm \bm{x}}{\mathrm{d}\tau} = \frac{1}{\mathpzc{D}(a)} \frac{\drm \bm{x}}{\drm a}.
\end{equation}
The Drift coefficient $\mathpzc{D}(a)$ is defined as
\begin{equation}
\mathpzc{D}(a) \equiv \frac{1}{a^2 \mathcal{H}(a)},
\end{equation}
with $\mathcal{H}(a) \equiv \drm a / (a \drm \tau)$ being the conformal expansion rate.

Second-order Lagrangian Perturbation Theory \citep[2LPT,][]{Bouchet1995} is built upon two potentials. The first-order potential $\phi^{(1)}$ is sourced by the initial overdensity field $\delta_\mathrm{i}$, while the second-order potential $\phi^{(2)}$ is sourced by the initial gravitational tidal field:
\begin{align}
\Delta_{\bm{q}} \phi^{(1)}(\bm{q}) &= \delta_\mathrm{i}(\bm{q}), \label{eq:potLPT1} \\
\Delta_{\bm{q}} \phi^{(2)}(\bm{q}) &= \sum_{\ell>m} \left[ \frac{\partial^2 \phi^{(1)}}{\partial \bm{q}_{\ell}^2} \frac{\partial^2 \phi^{(1)}}{\partial \bm{q}_m^2} - \left( \frac{\partial^2 \phi^{(1)}}{\partial \bm{q}_\ell \partial \bm{q}_m} \right)^2 \right]. \label{eq:potLPT2}
\end{align}
From these potentials, the time-independent LPT displacement fields $\boldsymbol{\varPsi}_1$ and $\boldsymbol{\varPsi}_2$ are defined as $\boldsymbol{\varPsi}_i \equiv \boldsymbol{\nabla}_{\bm{q}} \phi^{(i)}(\bm{q})$ for $i=1,2$. The 2LPT positions and momenta are then given by
\begin{align}
\bm{x}_{\mathrm{LPT}}(a) &\equiv \bm{q} - D_1(a) \, \boldsymbol{\varPsi}_1 + D_2(a) \, \boldsymbol{\varPsi}_2, \label{eq:2LPT_pos} \\
\mathpzc{D}(a) \,\bm{p}_{\mathrm{LPT}}(a) &\equiv -\frac{\mathrm{d}D_1}{\mathrm{d}a} \, \boldsymbol{\varPsi}_1 + \frac{\mathrm{d}D_2}{\mathrm{d}a} \, \boldsymbol{\varPsi}_2, \label{eq:2LPT_mom}
\end{align}
where $D_1$ and $D_2$ are the first- and second-order growth factors, respectively.

In the tCOLA framework, the full particle position of any particle is decomposed as \citep{Tassev2012b}
\begin{equation}
\bm{x}(a) = \bm{q} - D_1(a) \, \boldsymbol{\varPsi}_1 + D_2(a) \, \boldsymbol{\varPsi}_2 + \bm{x}_{\mathrm{res}}(a), \label{eq:residual_pos}
\end{equation}
where $\bm{x}_{\mathrm{res}}(a)$ is the residual displacement with respect to the 2LPT trajectory. We also define $\bm{p}_\mathrm{res}$ such that $\drm \bm{x}_\mathrm{res}/\drm a \equiv \mathpzc{D}(a) \, \bm{p}_\mathrm{res}$.
The time evolution of the system is governed by the Drift and Kick equations \citep{LeclercqThesis}:
\begin{align}
\deriv{\bm{x}}{a} &= \mathpzc{D}(a) \, \bm{p}_\mathrm{res}(a) - \deriv{D_1}{a} \, \boldsymbol{\varPsi}_1 + \deriv{D_2}{a} \, \boldsymbol{\varPsi}_2, \label{eq:drift} \\
\deriv{\bm{p}_\mathrm{res}}{a} &= \mathpzc{K}(a) \Bigl\lbrace \boldsymbol{\nabla}_{\bm{x}} \left( \Delta^{-1}_{\bm{x}} \delta \right)(a) - D_1(a) \boldsymbol{\varPsi}_1 \nonumber \\
&\quad + \left[ D_2(a) - D_1^2(a)\right] \boldsymbol{\varPsi}_2 \Bigr\rbrace, \label{eq:kick}
\end{align}
where the Kick coefficient $\mathpzc{K}(a)$ is defined by
\begin{equation}
\mathpzc{K}(a) \equiv -\frac{3}{2}\,\frac{\Omega_{\mathrm{m}}^{(0)} \left(\mathcal{H}^{(0)}\right)^2}{a\,\mathcal{H}(a)} ,
\end{equation}
and a superscript $(0)$ denotes quantities at the present time. The natural variables in tCOLA are $\bm{x}$ and $\bm{p}_\mathrm{res}$. The $L_-$ and $L_+$ operators perform the change of frame of reference by converting the physical momenta of particles to the residual momenta, and vice versa. They are defined by subtracting or adding the 2LPT momenta to the physical momenta, i.e.
\begin{equation}
\mathrm{L}_\pm(a) : \quad \bm{p}(a) \mapsto \bm{p}(a) \pm \bm{p}_\mathrm{LPT}(a).
\end{equation}

The continuous tCOLA equations \eqref{eq:drift} and \eqref{eq:kick} are discretised to allow integration over a finite number of time steps. We use $a$ as the time variable, and choose a symplectic Kick-Drift-Kick \citep[KDK,][]{Quinn1997} integrator, which alternates between Kick and Drift steps. In the following, the initial and final scale factors for a Drift step are denoted by $a_\mathrm{i}^\mathrm{D}$ and $a_\mathrm{f}^\mathrm{D}$, respectively, and by $a_\mathrm{i}^\mathrm{K}$ and $a_\mathrm{f}^\mathrm{K}$ for a Kick step. The COLA method usually employs a modified discretisation scheme, introduced by \citet{Tassev2013}, involving a free parameter $n_\mathrm{LPT}$ \citep[see][Appendix B]{Leclercq2020sCOLA}. Following earlier experimentations, we choose $n_\mathrm{LPT} = -2.5$, which best reproduces the results of full gravity.

Gravitational forces (the term $\boldsymbol{\nabla}_{\bm{x}} \left( \Delta^{-1}_{\bm{x}} \delta \right)$ in the Kick equation \eqref{eq:kick}) can, in principle, be evaluated using any numerical prescription. So far, implementations of tCOLA have relied on evaluating forces on a particle-mesh (PM) grid, which is the approach adopted here. To do so, we obtain the density field $\delta$ by mass assignment of particles onto a regular mesh using the Cloud-in-Cell \citep[CiC,][]{Hockney1981} method, from which the gravitational potential $\Phi = \Delta^{-1}_{\bm{x}} \delta$ is computed spectrally (see Section~\ref{ssec:Review of sCOLA}), and the force field $\boldsymbol{\nabla}_{\bm{x}} \Phi$ is derived using finite differences.

Since the residual $\bm{x}_\mathrm{res}$ with respect to the LPT trajectory is typically much smaller than the total displacement, fewer time steps are required to integrate equations \eqref{eq:drift} and \eqref{eq:kick} than the standard equations of motion to achieve results at the same accuracy. With respect to tree codes, tCOLA leads to a factor of ${\sim}100$ reduction in the number of time steps required to reach $1\%$ agreement down to scales of comoving wavenumber $k=0.5 \; h\,\mathrm{Mpc}^{-1}$ \citep{Tassev2013,Leclercq2015ST,Izard2016,Koda2016}.

    \subsection{Review of sCOLA}
    \label{ssec:Review of sCOLA}
    
        \begin{figure*}
          \includegraphics[width=\textwidth]{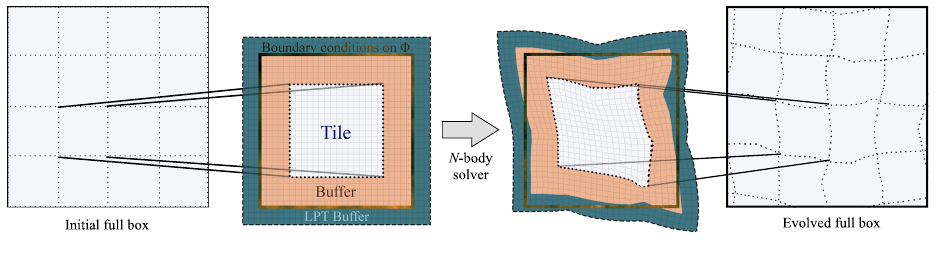}
          \caption{Illustration of sCOLA with an LPT buffer: an arbitrarily large volume is split into small tiles that are simulated independently. Particles are represented as vertices of the Lagrangian grid. The LPT buffer (new in this work, in blue) approximates the inflow of particles from outside the sCOLA box (defined as the tile in white, plus its buffer in orange) as the Lagrangian grid deforms. Non-periodic boundary conditions on the gravitational potential $\Phi$ (improved in this work) account for the large-scale gravitational field in which the sCOLA box is embedded.}
          \label{fig:sCOLA_concept}
        \end{figure*}

The spatial COmoving Lagrangian Acceleration framework \citep[sCOLA, originally introduced by][]{Tassev2015} extends COLA by using a comoving Lagrangian frame of reference to spatially decouple far and near fields in space, so that any single small volume embedded inside a larger simulation can reproduce the results of a standard $N$-body run. The idea was later extended and refined by \citet{Leclercq2020sCOLA} to simulate arbitrarily large volumes by splitting the simulation domain into small, independently evolved \emph{tiles}.
The key challenge is to make each tile self-contained: particles near the border of a tile are gravitationally influenced by particles outside it, and particles should be able to physically move in or out of the tile during the simulation. sCOLA addresses these two issues by introducing two approximations: (i) surrounding each tile with a \emph{buffer} region that pushes boundary artefacts away from the domain of interest, and (ii) imposing non-periodic boundary conditions on the gravitational potential $\Phi$ to encode the gravitational influence of the full matter distribution on each tile.
The result is a simulation scheme that is perfectly parallel and asynchronous: each tile can be evolved independently \citep{Leclercq2020sCOLA}.

We consider a full simulation volume of comoving size $L$, with $N_{\mathrm{p}}$ particles per axis and $N_{\mathrm{pm}}$ PM grid cells per axis, divided into $N_{\mathrm{tiles}}^3$ cubic tiles, each of size $L^{\mathrm{tile}}$ with $N_{\mathrm{p}}^{\mathrm{tile}} \equiv \frac{L^{\mathrm{tile}}}{L}N_{\mathrm{p}}$ particles per axis (left panel of Fig.~\ref{fig:sCOLA_concept}; see also Fig.~2 and Table~1 in \citealt{Leclercq2020sCOLA}).
Each tile is surrounded on all sides by a Lagrangian buffer of initial width $L^\mathrm{buffer}$. The Eulerian volume occupied by the tile and its buffer forms the \textit{sCOLA box}, which is evolved by the $N$-body solver. The generation of the initial conditions in sCOLA is slightly different from the traditional framework; indeed, the generation of the initial positions and momenta of particles in each sCOLA box can be included in the perfectly parallel section of the algorithm by directly computing them from a chunk of the 2LPT potentials $\phi^{(1)}$ and $\phi^{(2)}$ read locally (see Fig.~1 of \citealt{Leclercq2020sCOLA}).
This implies that memory-expensive operations needed on the full box (e.g.\ fast Fourier transforms) are limited to calculating the LPT potentials.

In a standard periodic PM simulation, the Poisson equation $\Delta_{\bm{x}} \Phi = \delta$ is solved globally via fast Fourier transforms, which assume periodic boundary conditions on the gravitational potential $\Phi$.
In contrast, in sCOLA, each sCOLA box is a finite sub-volume of a larger universe. Therefore, assuming periodicity would incorrectly wrap the gravitational influence of the sCOLA box on itself and be unphysical \citep[see][for a discussion of the resulting artefacts]{Tassev2015}.
Instead, following \citet{Leclercq2020sCOLA}, we impose Dirichlet boundary conditions on the faces of sCOLA boxes, prescribing the value $\Phi_\mathrm{BCs}$ of $\Phi$.
Solving the Poisson equation $\Delta_{\bm{x}} \Phi = \delta$ with Dirichlet boundary conditions requires the use of discrete sine transforms (DST) with ghost cells immediately outside the active domain of the PM grid \citep[][Appendix C]{Leclercq2020sCOLA}.

Boundary conditions $\Phi_\mathrm{BCs}$ should be as accurate as possible: errors in $\Phi_\mathrm{BCs}$ propagate inward over time, affecting the quality of the gravitational potential field throughout the box, particularly near the sCOLA box edges. Unfortunately, having access to the exact $\Phi_\mathrm{BCs}$ would require solving the simulation in the full box, which is what sCOLA seeks to avoid. Thus, to preserve the perfectly parallel nature of the algorithm, in \citet{Leclercq2020sCOLA}, the boundary conditions are approximated under the \emph{linearly-evolving potential} (LEP) approximation \citep{Brainerd1993,Bagla1994}:
\begin{equation}
\Phi_\mathrm{BCs}(\bm{x},a) \approx \Phi_\mathrm{LEP}(\bm{x},a) \equiv D_1(a)\,\phi^{(1)}(\bm{x}).
\end{equation}
In this approximation, the spatial structure of the potential is frozen to the initial potential linear field $\phi^{(1)} = \Phi_\mathrm{i}$ (whose computation is required for the initial conditions), with only its amplitude rescaled by the linear growth factor $D_1(a)$. All non-linear effects, such as gravitational clustering, are neglected.
 
    \subsection{Improvements of sCOLA}
    \label{ssec:Improvements of sCOLA}

In this work, we introduce two elements that improve the accuracy of the sCOLA algorithm, as described by \citet{Leclercq2020sCOLA}: (1) a LPT buffer to approximate the inflow of particles into sCOLA boxes, and (2) second-order boundary conditions on the gravitational potential.

\paragraph{LPT buffer}

As the simulation advances in time, particles that were initially outside the sCOLA box should be able to cross its boundary and enter the corresponding Eulerian volume.
The standard sCOLA approach considers particles only within the initial sCOLA box and therefore misses this inflow, producing an approximate interior density contrast $\delta_\mathrm{int}$ that sources the Poisson equation. This approximation produces artefacts near the boundary of sCOLA boxes and affects modes larger than the sCOLA box in the full simulation.
Of the two approximations involved in sCOLA ($\delta \approx \delta_\mathrm{int}$ and $\Phi_\mathrm{BCs} \approx \Phi_\mathrm{LEP}$), \citet[][Section~IV.C and Fig.~7]{Leclercq2020sCOLA} found this one to be the dominant one.
To address this shortcoming, we introduce for each tile a \emph{LPT buffer}, defined as a few additional Lagrangian layers of particles
surrounding the sCOLA box (see Fig.~\ref{fig:sCOLA_concept}). 
Within the LPT buffer, particles evolve solely under their 2LPT trajectories, i.e.\ no gravitational forces are computed for them.
They provide an inexpensive but physically motivated proxy for the inflow of matter from outside the sCOLA box and improve the approximation that defines $\delta_\mathrm{int}$ (see Appendix~\ref{app:LPT_buffer} for more details).
The LPT displacements $\boldsymbol{\varPsi}_1$ and $\boldsymbol{\varPsi}_2$ of these particles are generated from the 2LPT potentials $\phi^{(1)}$ and $\phi^{(2)}$ when the sCOLA box is initialised, within the perfectly parallel section of the algorithm.

\paragraph{Second-order boundary conditions}

After introducing the LPT buffer, our tests showed that the LEP approximation at the boundary of sCOLA boxes ($\Phi_\mathrm{BCs} \approx \Phi_\mathrm{LEP}$) became dominant over the approximation of the density contrast ($\delta \approx \delta_\mathrm{int}$). For this reason, in this work, we also replace the LEP approximation with an expression derived from second-order standard perturbation theory for the boundary conditions $\Phi_\mathrm{BCs}$. The new approximation, derived in Appendix~\ref{app:Phi2}, reads:
\begin{equation}
\Phi_{\mathrm{BCs}}(\bm{x},a) = D_1(a)\,\phi^{(1)}(\bm{x}) + D_1(a)^2\,\Phi_2(\bm{x}) ,
\end{equation}
where the second-order correction is:
\begin{equation}
\Phi_2 \equiv \frac{10}{7}\phi^{(2)} + \frac{1}{2}\Vert\boldsymbol{\varPsi}_1\Vert^2.
\end{equation}
This expression involves the second-order LPT potential $\phi^{(2)}$ and the squared norm of the first-order displacement $\boldsymbol{\varPsi}_1$, both quadratic in $\phi^{(1)}$.
Computation of $\Phi_{\mathrm{BCs}}(\bm{x},a)$ is inexpensive, since $\phi^{(2)}$ and $\boldsymbol{\varPsi}_1$ are computed for each particle of the sCOLA box upon initialisation.
Unlike the LEP approximation, the new $\Phi_\mathrm{BCs}(\bm{x},a)$ has a time-evolving spatial structure:
the second-order term captures some of the effects of non-linear structure formation on the gravitational potential.

Taken together, the two improvements described above increase the accuracy of the sCOLA algorithm (see Appendix~\ref{app:improvements} for a detailed study).
Therefore, to reach a given accuracy, they allow for smaller buffer regions around sCOLA tiles than reported by \citet{Leclercq2020sCOLA}: typically $L^{\mathrm{buffer}} \gtrsim 16\,h^{-1}\,\mathrm{Mpc}$ instead of $\gtrsim 40\,h^{-1}\,\mathrm{Mpc}$, see Appendix \ref{app:sCOLA_config}.
        
    \subsection{Lightcone building}
    \label{ssec:Lightcone building}
 
         \begin{figure}
             \centering
             \includegraphics[width=\linewidth]{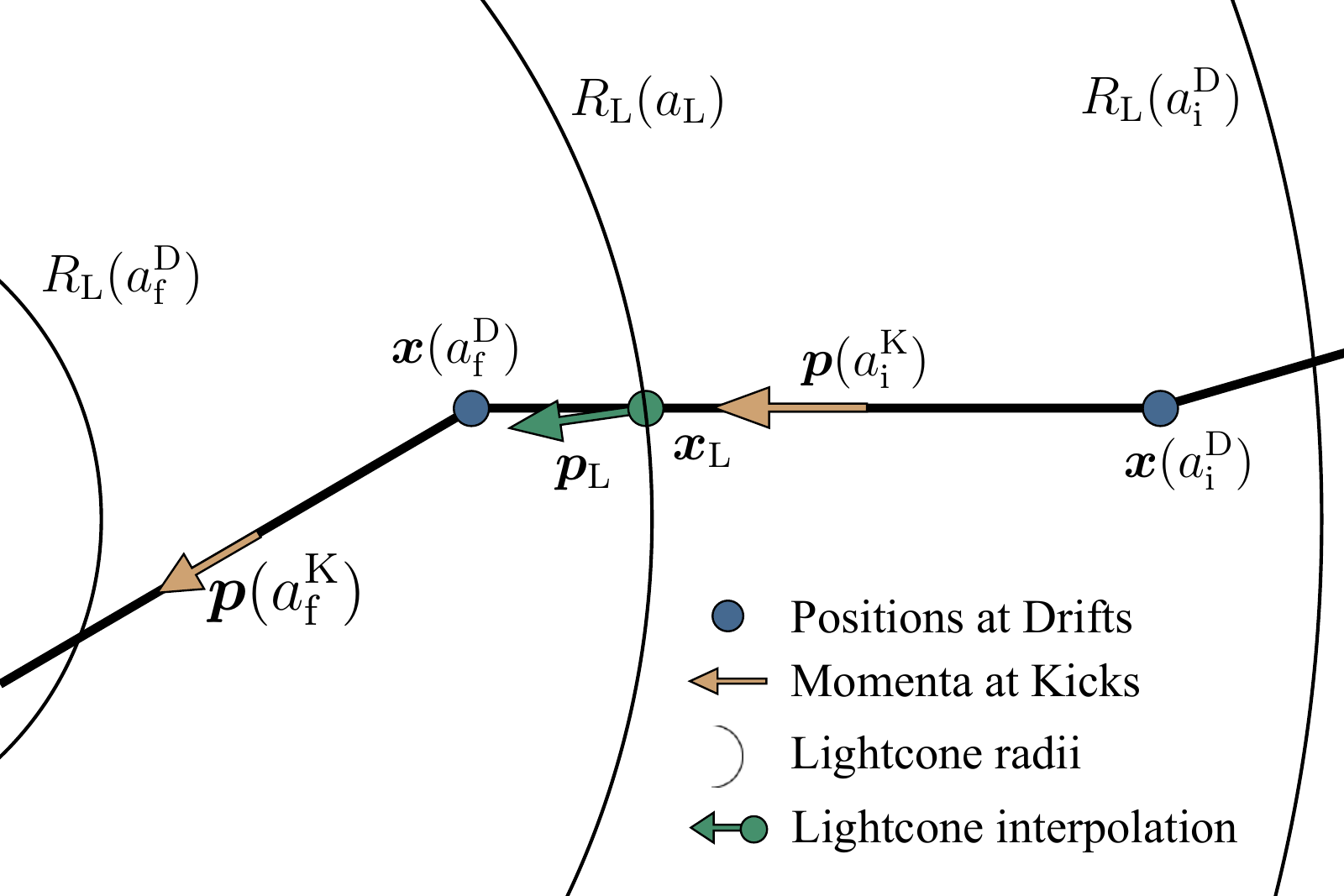}
             \caption{Interpolation of positions and momenta at lightcone crossing. The position $\bm{x}$ is interpolated between Drift times $a_\mathrm{i}^\mathrm{D}$ and $a_\mathrm{f}^\mathrm{D}$, at the lightcone crossing time $a_\mathrm{L}$ defined such that the particle lies at the comoving distance $R_\mathrm{L}(a_\mathrm{L})$ from the observer. The residual momentum $\bm{p}_\mathrm{res}$ is interpolated between the Kick times $a_\mathrm{i}^\mathrm{K}$ and $a_\mathrm{f}^\mathrm{K}$ that surround $a_\mathrm{L}$, and the LPT momentum at $a_\mathrm{L}$ is then added back to recover the full momentum $\bm{p}$.}
             \label{fig:lightcone_interp}
         \end{figure}

In a simulation, the observer sees its past lightcone, which is defined by the ensemble of null geodesics intersecting both the source and the observer's worldlines. An exact treatment of lightcone building using a relativistic ray-tracing framework is provided by \textsc{Magrathea-Pathfinder} \citep{MagratheaPathfinder1,MagratheaPathfinder2}.
Following previous work \citep[e.g.][]{LPICOLA,EuclidMock}, we use a simplified geodesic equation allowing us to construct the lightcone on-the-fly during the simulation, rather than in post-processing. We define the radius of the lightcone from the observer at any past scale factor $a$ as the comoving distance $R_\mathrm{L}(a)$ \citep{Hogg99}:
\begin{equation}
R_\mathrm{L}(a) \equiv \mathrm{c}\int_a^{1}\frac{\drm a'}{a' \, \mathcal{H}(a')},
\end{equation}
with $\mathcal{H}(a)$ the conformal expansion rate.

Let us consider a particle at position $\bm{x}_\mathrm{p}(a)$, which is at a comoving distance $R_\mathrm{p}(a) \equiv \Vert\bm{x}_\mathrm{p}(a) - \bm{x}_\mathrm{obs}\Vert$ of the observer located at $\bm{x}_\mathrm{obs}$. It will be seen by the observer when it crosses its past lightcone, i.e.\ at a time $a_\mathrm{L}$ such that
\begin{equation}
R_\mathrm{L}(a_\mathrm{L}) = R_\mathrm{p}(a_\mathrm{L}).
\end{equation}
This equation is the \emph{lightcone crossing condition}.
Defining $\Delta R \equiv R_\mathrm{L} - R_\mathrm{p}$, the crossing condition reads $\Delta R(a_\mathrm{L}) = 0$. We assume that $\Delta R$ is a continuous function of $a$. Since particles move slower than light, the particle distance $R_\mathrm{p}(a)$ cannot decrease faster than the horizon $R_\mathrm{L}(a)$ as $a$ increases, so $\Delta R(a)$ is a decreasing function of $a$ and $\Delta R(1) \leq 0$ (because $R_\mathrm{L}(1) = 0$). For observable particles (particles within the observable radius $R_\mathrm{L}(0)$), we have $\Delta R(0) \geq 0$. Therefore, for all observable particles, there exists a scale factor $a_\mathrm{L}$ such that $\Delta R(a_\mathrm{L}) = 0$.

Since particle trajectories are integrated in discrete Drift steps, positions $\bm{x}$ are only defined at discrete scale factors $a_\mathrm{i}^\mathrm{D}, a_\mathrm{f}^\mathrm{D}$. A lightcone crossing occurs between $a_\mathrm{i}^\mathrm{D}$ and $a_\mathrm{f}^\mathrm{D}$ when:
\begin{equation}
\Delta R(a_\mathrm{i}^\mathrm{D}) > 0 \qquad \text{and} \qquad \Delta R(a_\mathrm{f}^\mathrm{D}) \leq 0.
\end{equation}
Thus, at each time step, for all particles that have not yet crossed the lightcone, we monitor the sign of $\Delta R$. When it becomes non-positive, the particle is flagged, and its position $\bm{x}_\mathrm{L}$ and residual momentum $\bm{p}_\mathrm{res,L}$ at $a_\mathrm{L}$ are recovered by interpolation. For periodic simulations, the evaluation of $\Delta R(a_\mathrm{f}^\mathrm{D})$ and interpolation of $\bm{x}_\mathrm{L}$ are done before wrapping particle positions across the box boundaries to avoid unphysical discontinuities in the interpolation.

For a flagged particle, the crossing time $a_\mathrm{L}$ and position $\bm{x}_\mathrm{L}$ are obtained by linear interpolation between $a_\mathrm{i}^\mathrm{D}$ and $a_\mathrm{f}^\mathrm{D}$: using $\Delta R(a_\mathrm{L}) = 0$,
\begin{align}
\alpha &\equiv \frac{-\Delta R(a_\mathrm{i}^\mathrm{D})}{\Delta R(a_\mathrm{f}^\mathrm{D}) - \Delta R(a_\mathrm{i}^\mathrm{D})}, \\
a_\mathrm{L} &\approx a_\mathrm{i}^\mathrm{D} + \alpha \left(a_\mathrm{f}^\mathrm{D} - a_\mathrm{i}^\mathrm{D}\right), \\
\bm{x}_\mathrm{L} &\approx \bm{x}(a_\mathrm{i}^\mathrm{D}) + \alpha \left[\bm{x}(a_\mathrm{f}^\mathrm{D}) - \bm{x}(a_\mathrm{i}^\mathrm{D})\right].
\end{align}
The residual momenta $\bm{p}_\mathrm{res}$ are defined at Kick times. Therefore, the residual momentum at lightcone crossing $\bm{p}_\mathrm{res,L}$ is interpolated between the two Kick times $a_\mathrm{i}^\mathrm{K}$ and $a_\mathrm{f}^\mathrm{K}$ bracketing $a_\mathrm{L}$, i.e.\ such that $a_\mathrm{i}^\mathrm{K} < a_\mathrm{L} \leq a_\mathrm{f}^\mathrm{K}$:
\begin{equation}
\bm{p}_\mathrm{res,L} \approx \bm{p}_{\mathrm{res}}(a_\mathrm{i}^\mathrm{K}) + \frac{a_\mathrm{L} - a_\mathrm{i}^\mathrm{K}}{a_\mathrm{f}^\mathrm{K} - a_\mathrm{i}^\mathrm{K}} \left[\bm{p}_{\mathrm{res}}(a_\mathrm{f}^\mathrm{K}) - \bm{p}_{\mathrm{res}}(a_\mathrm{i}^\mathrm{K})\right].
\end{equation}
The LPT momentum $\bm{p}_\mathrm{LPT}$ at $a_\mathrm{L}$ (from equation~\eqref{eq:2LPT_mom}) is then added back via the $L_+$ operator to recover the full momentum $\bm{p}$.
Figure~\ref{fig:lightcone_interp} summarises the above interpolation procedure.

        \begin{figure}
            \centering
            \includegraphics[width=\linewidth]{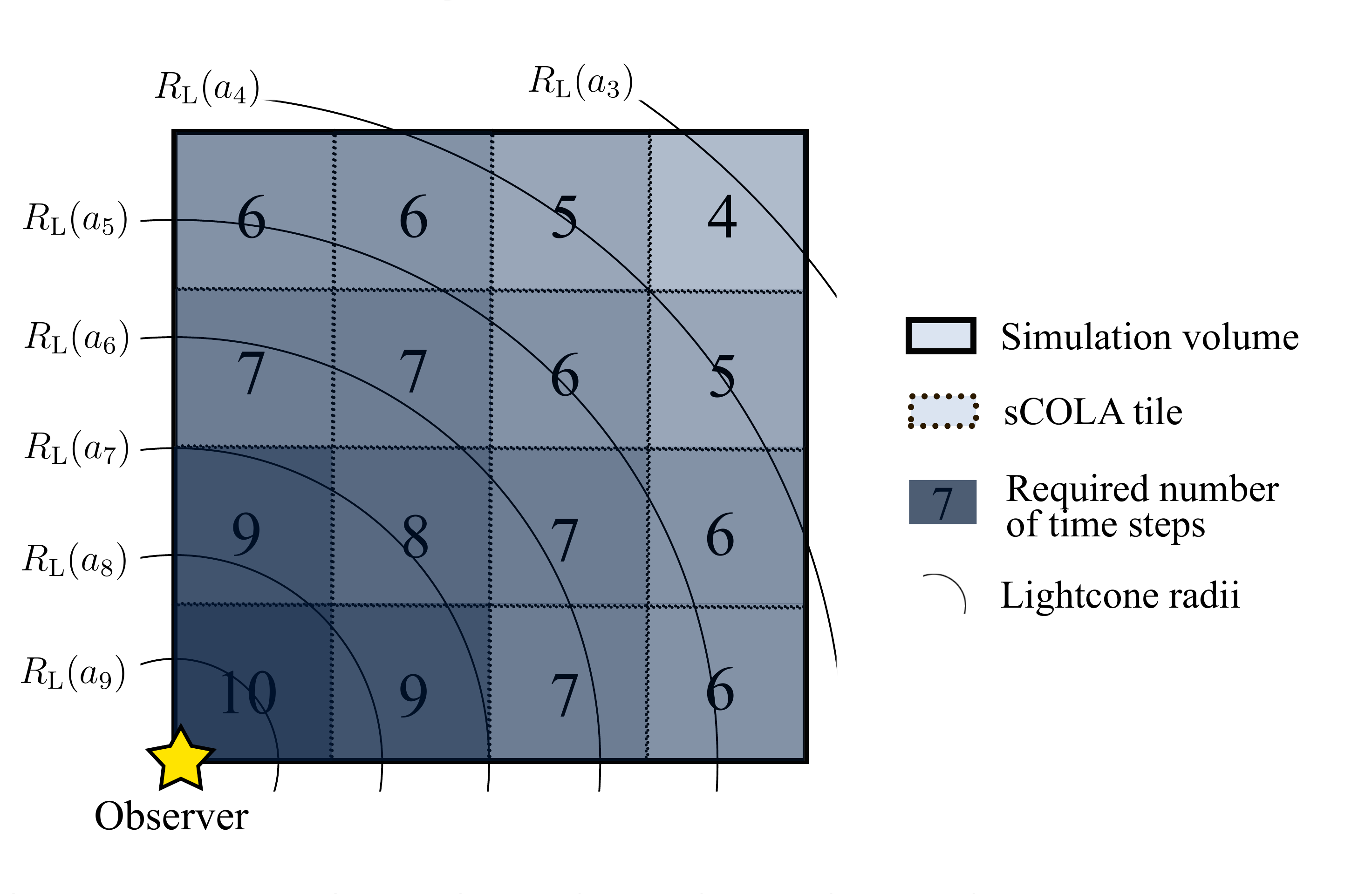}
            \caption{sCOLA lightcone building: each tile is evolved only until the lightcone has left it, reducing the number of required time steps. Tiles farther from the observer are crossed by the lightcone earlier and require fewer time steps.}
            \label{fig:lightcone_scola}
        \end{figure}

\begin{figure}
            \centering
            \includegraphics[width=\linewidth]{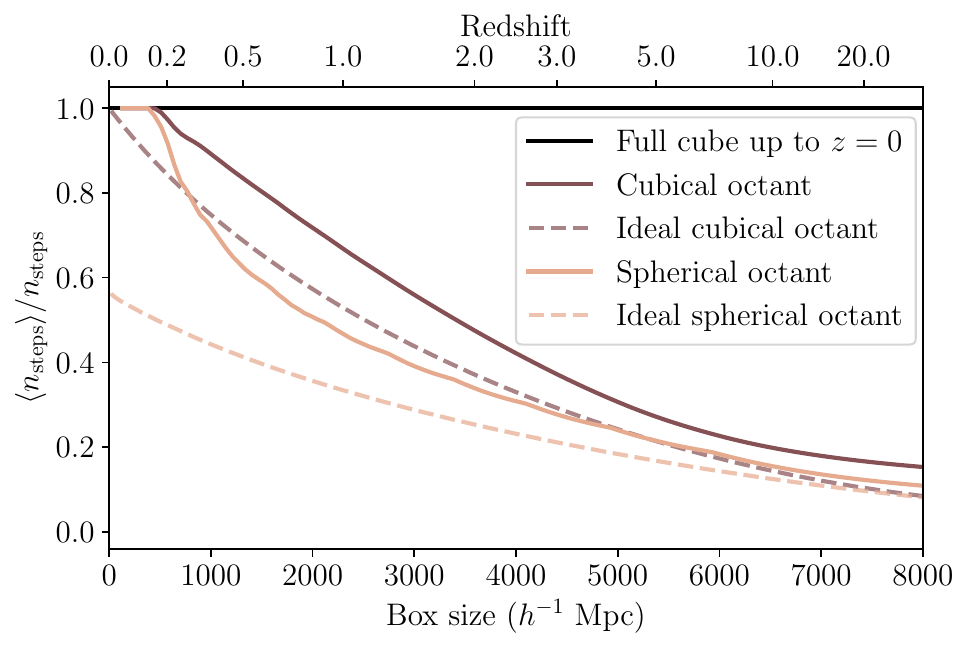}
            \caption{Mean number of time steps needed to build a lightcone with sCOLA, $\langle n_\mathrm{steps}\rangle$ (averaged over sCOLA boxes), divided by the number of time steps $n_\mathrm{steps}$ needed to evolve each sCOLA box to $a_0=1$, as a function of the size of the full box. We distinguish the cases of a cubical octant (the full volume is evolved) and a spherical octant (only the tiles in the sphere are evolved). We compare the ideal case of arbitrarily small tiles and an infinite number of time steps (dashed lines) to the realistic configuration used in this work, i.e.\ ten time steps and tiles of size $64\,h^{-1}\,\mathrm{Mpc}$ (solid lines). For the $3.84\,h^{-1}\,\mathrm{Gpc}$ cubical octant of this work, building the lightcone with sCOLA requires 57\% fewer time steps than evolving the full volume to $a_0=1$.}
            \label{fig:fraction_tiles}
        \end{figure}

For a monolithic tCOLA simulation with the observer inside the box, the full simulation is run from the initial time to $a_0 = 1$, and lightcone outputs $\bm{x}_\mathrm{L}$, $\bm{p}_\mathrm{L}$ are collected progressively as particles cross the lightcone.
In the case of sCOLA, the simulation of each box can be stopped as soon as the lightcone has left the sCOLA box. Indeed, once the lightcone has passed the corner of the box closest to the observer, all particles in the tile have already crossed the lightcone and their outputs $\bm{x}_\mathrm{L}$, $\bm{p}_\mathrm{L}$ have been collected. Tiles farther from the observer are crossed by the lightcone at earlier times (smaller $a$), so they require fewer time steps, as illustrated in Fig.~\ref{fig:lightcone_scola}.
In Figure~\ref{fig:fraction_tiles}, we show the average fraction of time steps needed as a function of the size of the box. For example, sCOLA lightcone building for a $3.84\,h^{-1}\,\mathrm{Gpc}$ octant with 10 time steps linearly spaced in $D_1$ yields a 57\% reduction in the average number of time steps with respect to evolving all of the sCOLA boxes to $a_0 = 1$.
 
\texttt{Gadget4}, monolithic tCOLA, and sCOLA handle the periodic wrapping of particles differently when building a lightcone, leading to differences near the edges. \texttt{Gadget4} replicates the box, which means that there can be duplication of particles in the produced snapshot. Monolithic tCOLA wraps particles during their evolution, so particles can cross the full box boundary and be wrapped before being interpolated. Finally, sCOLA does not wrap particles during evolution but does so at the end when gathering the tiles into a single snapshot. Due to these differences, we remove a safe border of $30\,h^{-1}\,\mathrm{Mpc}$ around the box edges before any comparative analysis.

\section{Simulations}
\label{sec:Simulations}

In Section~\ref{ssec:General setup}, we describe the two simulation volumes used throughout this work and the generation of initial conditions. In Section~\ref{ssec:Reference with Gadget4}, we present the reference \texttt{Gadget4} run; in Section~\ref{ssec:Monolithic tCOLA}, the monolithic tCOLA run, which isolates the error due to the sCOLA tiling from that due to the COLA time-stepping; and in Section~\ref{ssec:sCOLA}, the sCOLA runs themselves.

    \subsection{General setup}
    \label{ssec:General setup}
    
    \begin{table*}[]
    \centering
    
    \newlength{\cw}
    \setlength{\cw}{0.6cm}  
    
    \resizebox{\textwidth}{!}{%
        \begin{tabular}{c*{22}{>{\centering\arraybackslash}p{\cw}}}
            \hline\hline
            Kicks
              & \multicolumn{1}{p{\cw}|}{0.05}
              & \multicolumn{2}{p{2\cw}|}{0.0868}
              & \multicolumn{2}{p{2\cw}|}{0.1606}
              & \multicolumn{2}{p{2\cw}|}{0.2351}
              & \multicolumn{2}{p{2\cw}|}{0.3111}
              & \multicolumn{2}{p{2\cw}|}{0.3897}
              & \multicolumn{2}{p{2\cw}|}{0.4727}
              & \multicolumn{2}{p{2\cw}|}{0.5623}
              & \multicolumn{2}{p{2\cw}|}{0.6620}
              & \multicolumn{2}{p{2\cw}|}{0.7771}
              & \multicolumn{2}{p{2\cw}|}{0.9165}
              & \multicolumn{1}{p{\cw}}{1.00}
            \\
            \hline
            Drifts
              & \multicolumn{2}{p{2\cw}|}{0.0500}
              & \multicolumn{2}{p{2\cw}|}{0.1236}
              & \multicolumn{2}{p{2\cw}|}{0.1977}
              & \multicolumn{2}{p{2\cw}|}{0.2728}
              & \multicolumn{2}{p{2\cw}|}{0.3500}
              & \multicolumn{2}{p{2\cw}|}{0.4305}
              & \multicolumn{2}{p{2\cw}|}{0.5165}
              & \multicolumn{2}{p{2\cw}|}{0.6106}
              & \multicolumn{2}{p{2\cw}|}{0.7171}
              & \multicolumn{2}{p{2\cw}|}{0.8430}
              & \multicolumn{2}{p{2\cw}}{1.000}
            \\
            \hline\hline
        \end{tabular}%
    }
    \caption{Scale factors of the Kick and Drift operations of the time-stepping used in this article. Each operation runs between two consecutive scale factors (for instance the first Kick is between $0.0500$ and $0.0868$).}
    \label{tab:time_stepping}
    \end{table*}

We use two simulation setups, \textit{Main} and \textit{Mini}. The \textit{Main} simulations have a box side length $L=3.84\,h^{-1}\,\mathrm{Gpc}$, $N_\mathrm{p}=3\,840$ particles per dimension, and $N_\mathrm{pm}=2N_\mathrm{p}=7\,680$ PM grid cells per dimension. The particle mass is $m_\mathrm{p} = 8.62\times10^{10}\,h^{-1}\,\mathrm{M}_\odot$. The box is slightly larger than that of the \textit{Euclid Flagship} simulation \citep{EuclidMock} ($L^\mathrm{Euclid~Flagship}=3.60\, h^{-1}\,\mathrm{Gpc}$) but the resolution is coarser
($N_\mathrm{p}^\mathrm{Euclid~Flagship}=16\,000$).
To compare sCOLA to monolithic tCOLA, we also ran \textit{Mini} simulations, with the same mass and spatial resolution as the \textit{Main} simulations but half their box side length: $L^\mathrm{Mini}=1.92\, h^{-1}\,\mathrm{Gpc}$, $N_\mathrm{p}^\mathrm{Mini}=1\,920$, $N_\mathrm{pm}^\mathrm{Mini}=3\,840$.

The initial conditions were generated with the \texttt{MUSIC2-monofonIC} code \citep{monofonIC}, modified to write to disk the quantities required by \texttt{Simbelmynë} and \texttt{Gadget4}: the initial overdensity field $\delta_\mathrm{i}$, the 2LPT potentials $\phi^{(1)}$ and $\phi^{(2)}$, and the Lagrangian positions $\bm{q}$. \texttt{monofonIC} relies on \texttt{CLASS} \citep{CLASS} for the initial power spectrum. The initial conditions were generated at scale factor $a_{\mathrm{i}}=0.05$. The cosmological parameters are those of \citet{Planck2018VI}: $\Omega_\mathrm{m}=0.311$, $\Omega_\mathrm{b}=0.049$, $\Omega_\Lambda=0.689$, $H_0=67.7\,\mathrm{km}\,\mathrm{s}^{-1}\,\mathrm{Mpc}^{-1}$, $n_\mathrm{s}=0.967$, and $A_\mathrm{s}=2.105\times10^{-9}$.

For both tCOLA and sCOLA, we used the KDK leapfrog integrator, with 10 time steps evenly spaced in $D_1(a)$ \citep[following][]{Bullfrog}. Table~\ref{tab:time_stepping} lists the scale factors of the corresponding Kick and Drift operations.

    \subsection{Reference with Gadget4}
    \label{ssec:Reference with Gadget4}

We used \texttt{Gadget4} \citep{Gadget4} as the reference $N$-body code against which we assess the accuracy and computational cost of sCOLA. It was started from the particles generated by \texttt{MUSIC2-monofonIC} at $a_{\mathrm{i}}=0.05$.
We used the TreePM force computation of \texttt{Gadget4} with a $3\,840^3$ PM grid (\texttt{PMGRID}), a force-split scale of $1.25$ mesh cells (\texttt{ASMTH}), and a comoving softening length of $50\,h^{-1}\,\mathrm{kpc}$ (1/20 of the mean inter-particle separation); the tree was opened with the relative criterion at $\texttt{ErrTolForceAcc}=0.005$, the time-step accuracy parameter was $\texttt{ErrTolIntAccuracy}=0.025$, and long- and short-range forces were integrated on a common time step (\texttt{TREEPM\_NOTIMESPLIT}).
We ran \texttt{Gadget4} on the \texttt{Irene} cluster of the Très Grand Centre de calcul du CEA (TGCC). The \textit{Main} simulation was run on 192 nodes (128 CPUs and 228 GB of RAM per node), in many short jobs of about one hour each. In total, it consumed $3\,379\,000$ CPU hours (CPUh).
We also ran a \textit{Mini} reference simulation with \texttt{Gadget4}, using the same parameters except for the PM grid, set to $1\,920^3$; it consumed $128\,200$ CPUh.

    \subsection{Monolithic tCOLA}
    \label{ssec:Monolithic tCOLA}

To separate the error due to the sCOLA approximations from that due to the reduced number of COLA time steps, we ran a monolithic tCOLA simulation with the \texttt{Simbelmynë} code \citep{Leclercq2015ST,Leclercq2020sCOLA}. We used an upgraded version of the code, which includes the sCOLA improvements and the lightcone-building capability described in Section~\ref{sec:Methods}. 
Since the code does not support multi-node computation, the number of particles in a monolithic run is limited by the memory of a single node ($1.02\times10^{10}$ particles for a 2 TB node). We therefore ran tCOLA only in the \textit{Mini} setup.
The tCOLA \textit{Mini} simulation was run on the \texttt{Infinity} cluster of the Institut d'Astrophysique de Paris (IAP), on a single node with 128 CPUs and 2 TB of RAM, for 1~h~36~min of wall-clock time (205 CPUh).
To disentangle the effects of the time integration from those of the force model, we also ran a tCOLA \textit{Mini} simulation with $100$ time steps ($1\,527$ CPUh).

    \subsection{sCOLA}
    \label{ssec:sCOLA}

For the \textit{Main} sCOLA simulations, we used $N_\mathrm{tiles}=60$ tiles per dimension, for a total of $216\,000$ tiles; and for the \textit{Mini} sCOLA simulations, $N_\mathrm{tiles}=30$, for a total of $27\,000$ tiles. Each tile thus had a side length $L^\mathrm{tile}=64\,h^{-1}\,\mathrm{Mpc}$ and $N_\mathrm{p}^\mathrm{tile}=64$ particles per dimension.
We used buffers of width $L^\mathrm{buffer}=16\,h^{-1}\,\mathrm{Mpc}$ ($N_\mathrm{p}^\mathrm{buffer}=16$). Each sCOLA box is evolved in full, but only the particles of its tile are kept; the number of particles evolved therefore exceeds the number kept by the oversimulation factor \citep{Leclercq2020sCOLA}
\begin{equation}
r \equiv \frac{\left(L^\mathrm{tile}+2L^\mathrm{buffer}\right)^3}{\left(L^\mathrm{tile}\right)^3} = \left(1+\frac{2L^\mathrm{buffer}}{L^\mathrm{tile}}\right)^3 ,
\label{eq:oversimulation}
\end{equation}
which equals 3.38 for the configuration used here. Around the buffer, we added an LPT buffer of width $L^\mathrm{LPT~buffer}=8\,h^{-1}\,\mathrm{Mpc}$ ($N_\mathrm{p}^\mathrm{LPT~buffer}=8$), in which particles move only under their 2LPT displacements.
Appendix~\ref{app:sCOLA_config} justifies the choice of these sCOLA parameters.

We ran the \textit{Main} simulation with sCOLA twice: once up to $a_0=1$, and a second time with lightcone-only output, in which case each sCOLA box is stopped as soon as the lightcone has left it (Section~\ref{ssec:Lightcone building}). The sCOLA \textit{Main} and \textit{Mini} simulations were run asynchronously on the \texttt{Infinity} cluster of the IAP, on whichever CPUs were available at the time. Each sCOLA box used 4 CPUs and 150 MB of RAM.
We also ran sCOLA simulations with $100$ time steps for both \textit{Mini} and \textit{Main}, and both up to $a_0=1$ and lightcone-only.

    \begin{figure*}[t]
        \centering
        \includegraphics[width=\textwidth]{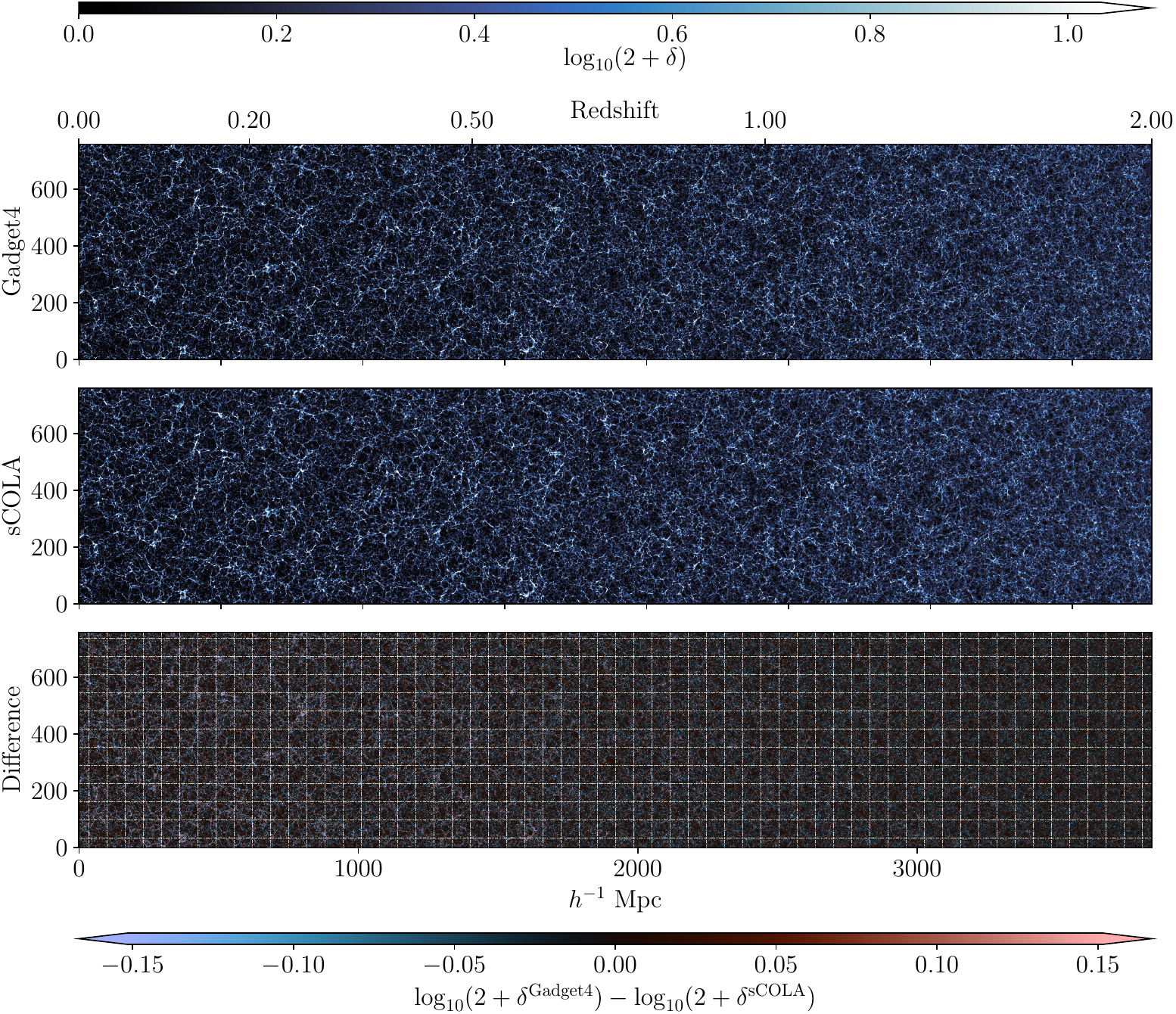}
        \caption{Slice of the \textit{Main} simulation lightcone dark matter density field for \texttt{Gadget4} (top row) and sCOLA (middle row), shown as $\log_{10}(2+\delta)$. The bottom row displays the difference in $\log_{10}(2+\delta)$ between the two fields, with the sCOLA tiling represented as dashed lines. The observer is located at $(0,0)$ (bottom left corner), coordinates are in $h^{-1}\,\mathrm{Mpc}$, and the full box is a cube of side $3.84\,h^{-1}\,\mathrm{Gpc}$. sCOLA reproduces the \texttt{Gadget4} density field with no visible tile-boundary artefacts, with a mean absolute difference in $\log_{10}(2+\delta)$ of $0.036$.}
        \label{fig:density_lightcone}
    \end{figure*}
    
\section{Results}
\label{sec:Results}

We now compare our sCOLA lightcone to the reference \texttt{Gadget4} simulation, and assess the cost of producing it. Section~\ref{ssec:Density and velocity fields} inspects the density and velocity fields directly, where artefacts of the tiling would be most conspicuous. Section~\ref{ssec:Power spectra and bispectra} turns to summary statistics, which quantify the agreement scale by scale and allow the residual differences to be attributed to individual approximations. Section~\ref{ssec:Computational cost} reports the computational cost of the method, including the savings obtained by stopping each tile as soon as the lightcone has left it.

    \subsection{Density and velocity fields}
    \label{ssec:Density and velocity fields}
    
Figure~\ref{fig:density_lightcone} compares a slice of the \textit{Main} simulation lightcone density field from sCOLA and \texttt{Gadget4}. We show the quantity $\log_{10}(2+\delta)$ to enhance the visual contrast between voids and high-density regions. The mean absolute difference in $\log_{10}(2+\delta)$ between the two fields is $0.036$, and 
no discontinuity is visible at tile boundaries: the sCOLA tiling preserves the continuity of the density field despite the independent evolution of the tiles.
    
        \begin{figure*}
            \centering
            \includegraphics[width=\textwidth]{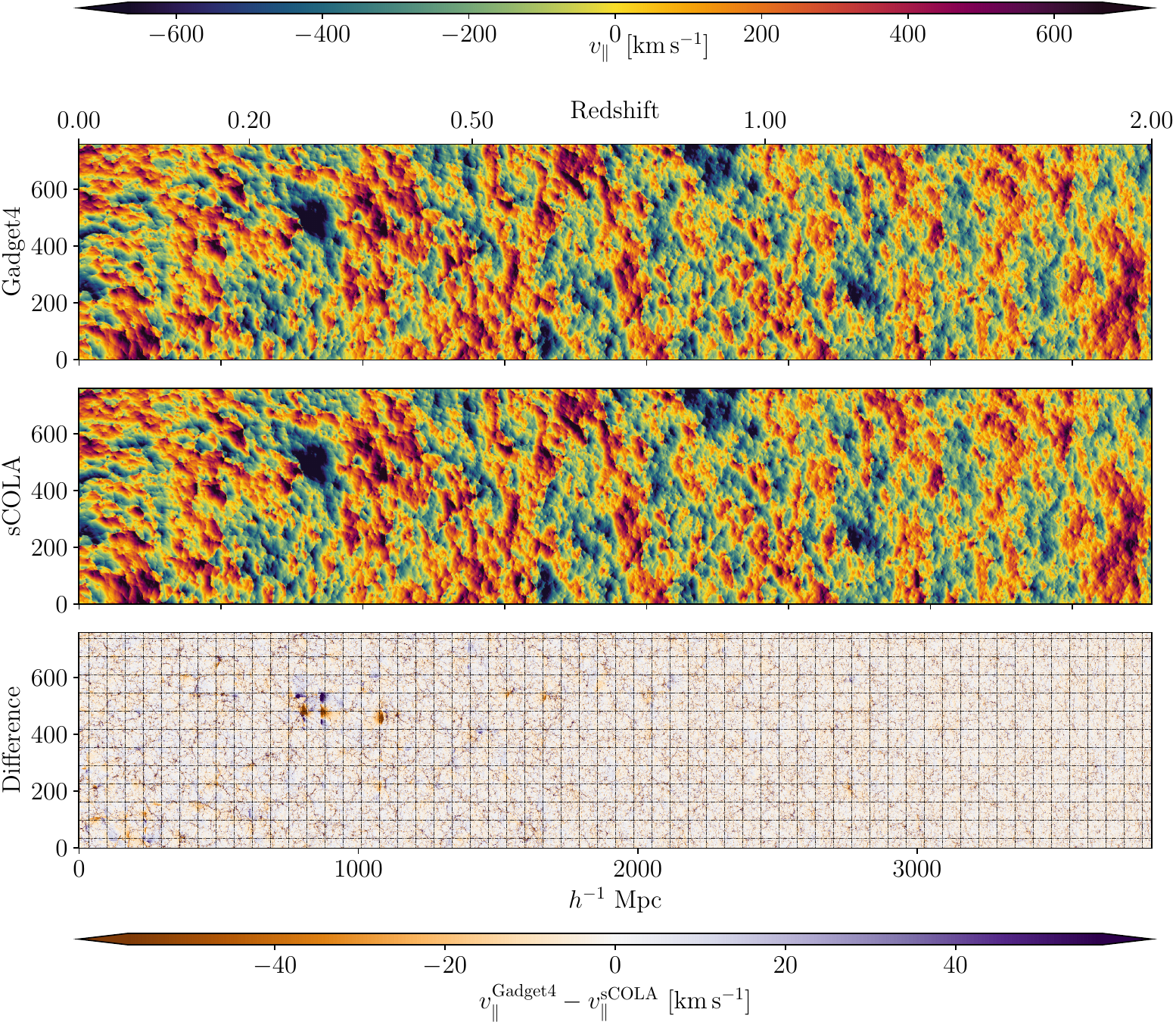}
            \caption{Slice of the \textit{Main} simulation lightcone velocity field for \texttt{Gadget4} (top row) and sCOLA (middle row). We plot the line-of-sight velocity component $v_\parallel$ in $\mathrm{km}\,\mathrm{s}^{-1}$. The bottom row displays the difference between the two velocity fields, with the sCOLA tiling represented as dashed lines.
            sCOLA reproduces the \texttt{Gadget4} velocity field with only a handful of visible tile-boundary artefacts; the mean absolute difference in $v_\parallel$ is $9.67\,\mathrm{km}\,\mathrm{s}^{-1}$.}
            \label{fig:velocity_lightcone}
        \end{figure*}
    
Figure~\ref{fig:velocity_lightcone} compares a $Y$-$Z$ slice of the line-of-sight velocity field from sCOLA and \texttt{Gadget4}. The velocity field was reconstructed using the Simplex-in-Cell (SiC) projection \citep{SiC1,SiC2,HahnAnguloAbel2015,Leclercq2017DMSHEET}, and it is shown as $v_\parallel$ in $\mathrm{km}\,\mathrm{s}^{-1}$. sCOLA recovers the large-scale velocity structure of \texttt{Gadget4}, with a mean absolute difference in $v_\parallel$ of $9.67\,\mathrm{km}\,\mathrm{s}^{-1}$.
A few tile-related velocity defects remain, for instance around $(800,480)~h^{-1}\,\mathrm{Mpc}$ in Fig.~\ref{fig:velocity_lightcone}. We traced them to particle displacements exceeding the buffer width, and checked that they disappear when the buffer is enlarged to $32\,h^{-1}\,\mathrm{Mpc}$.

    \subsection{Power spectra and bispectra}
    \label{ssec:Power spectra and bispectra}

        \begin{figure*}
            \centering
            \includegraphics[width=\textwidth]{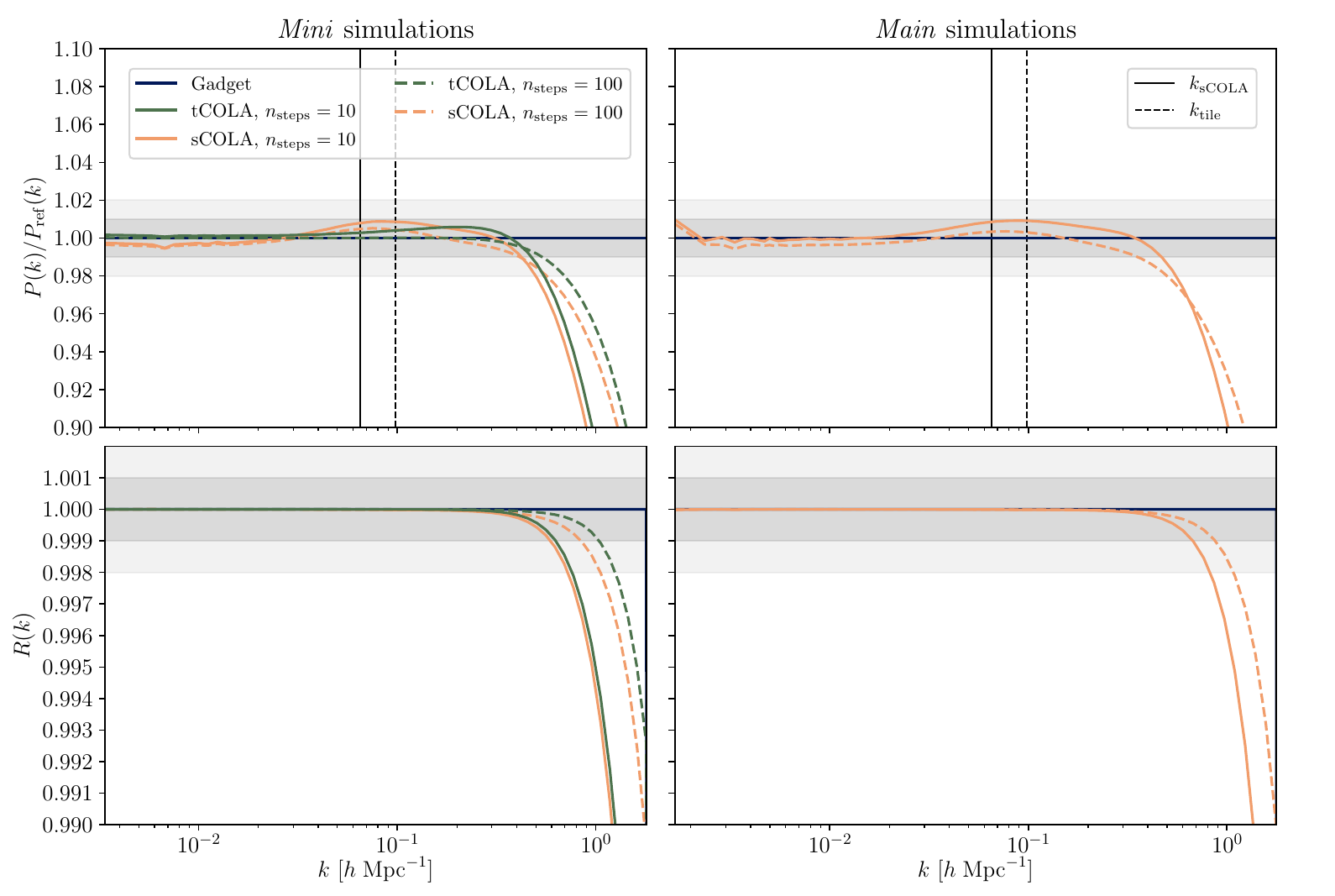}
            \caption{Comparison of the matter density power spectrum and cross-correlations between \texttt{Gadget4}, sCOLA and tCOLA on the \textit{Mini} and \textit{Main} lightcone simulations, for 10 and 100 time steps ($n_\mathrm{steps}$ variable). For the power spectrum, we plot the ratio to the reference \texttt{Gadget4} power spectrum. The grey bands are 1\% and 2\% around \texttt{Gadget4}. The cross-correlations are also computed with respect to \texttt{Gadget4}; there, the grey bands are 1\textperthousand\space and 2\textperthousand. sCOLA reaches 1\% agreement with \texttt{Gadget4} up to $k\simeq 0.4\,h\,\mathrm{Mpc}^{-1}$ with 10 time steps; the residual bump near the sCOLA box scale stays within 1\%. The fundamental modes of the tile and the sCOLA box are denoted $k_\mathrm{tile}$ and $k_\mathrm{sCOLA}$, respectively.
            }
            \label{fig:power_spectrum_Main}
        \end{figure*}
    
Figure~\ref{fig:power_spectrum_Main} compares two-point statistics in Fourier space for sCOLA, monolithic tCOLA, and \texttt{Gadget4}, allowing us to separate three sources of discrepancy:
(i) the reduced number of time steps (COLA),
(ii) the force model (PM versus TreePM),
and (iii) the spatial tiling (sCOLA).
With 10 time steps, sCOLA reaches $1\%$ agreement with \texttt{Gadget4} on the matter power spectrum up to $k \simeq 0.4\,h\,\mathrm{Mpc}^{-1}$, and cross-correlations above $0.999$ up to $k \simeq 0.6\,h\,\mathrm{Mpc}^{-1}$.
    
The sCOLA power spectrum shows a mild bump with respect to tCOLA near the fundamental mode of the sCOLA box ($k_\mathrm{sCOLA} \sim 0.065\,h\,\mathrm{Mpc}^{-1}$, i.e.\ $2\pi/L^\mathrm{sCOLA}$ with $L^\mathrm{sCOLA} = L^\mathrm{tile}+2L^\mathrm{buffer} = 96\,h^{-1}\,\mathrm{Mpc}$, the side of the sCOLA box and not of the tile), associated with the finite size of the sCOLA boxes.
For the chosen configuration, its amplitude remains below $1\%$. The bump can be further reduced by changing the sCOLA configuration (see Appendix~\ref{app:sCOLA_config}): with $L^\mathrm{tile}=192\,h^{-1}\,\mathrm{Mpc}$ and $L^\mathrm{buffer}=16\,h^{-1}\,\mathrm{Mpc}$, for instance, its amplitude falls below $0.2\%$.
    
Increasing the number of time steps from 10 to 100 in sCOLA and tCOLA reduces the overall difference from \texttt{Gadget4} and pushes the small-scale divergence to smaller scales. For sCOLA, the wavenumber up to which the cross-correlation exceeds $0.999$ moves from $k \simeq 0.6\,h\,\mathrm{Mpc}^{-1}$ to $k \simeq 0.9\,h\,\mathrm{Mpc}^{-1}$. The divergence at small scales is caused by the limited resolution of the PM force model, as opposed to the TreePM method of \texttt{Gadget4}.
    
We also computed the equilateral bispectrum (see Appendix~\ref{app:bispectra}). sCOLA agrees with \texttt{Gadget4} at the $\sim 2\%$ level up to $k \simeq 0.4\,h\,\mathrm{Mpc}^{-1}$, confirming that it captures the non-Gaussian features of the density field. The velocity divergence power spectrum agrees with \texttt{Gadget4} at the $1.5\%$ level up to $k \simeq 0.6\,h\,\mathrm{Mpc}^{-1}$
(see Appendix~\ref{app:velocity_ps}).
The power spectra and bispectra were computed on the lightcone, so the \textit{Main} and \textit{Mini} simulations probe different redshift ranges, $0\leq z\leq 10.9$ and $0\leq z \leq 1.75$ respectively; this difference in redshift coverage explains the difference between the \textit{Main} and \textit{Mini} spectra, notably at small scales.
    
    \subsection{Computational cost}
    \label{ssec:Computational cost}
    
        \begin{table*}[]
            \begin{tabular}{llllll}
            \hline \hline
            \textbf{Project}                                                             & \textbf{Gravity Model}      & $\bm{N_\textbf{p}}^3$              & $\bm{L}\;(h^{-1}\,\mathrm{Mpc})$ & $\bm{m_\mathrm{p}}\;(h^{-1}\,\mathrm{M}_\odot)$ & \begin{tabular}[c]{@{}l@{}}\textbf{CPU hours}\\ \textbf{ / million }$\bm{N_\textbf{p}}$ \end{tabular} \\ \hline
            \begin{tabular}[c]{@{}l@{}}Uchuu\\ \citep{Uchuu}\end{tabular}                & \texttt{GreeM}     & $12\,800^3$                          & $2\,000$                             & $3.27\times10^{8}$                             & $\sim$ 9.54                                                                               \\ \hline
            \begin{tabular}[c]{@{}l@{}}\textit{Euclid Flagship}\\ \citep{EuclidMock}\end{tabular} & \texttt{PKDGRAV3}  & $16\,000^3$                          & $3\,600$                             & $1.00\times10^{9}$                             & $\sim$ 16.6\footnote{The \textit{Euclid Flagship} was run on GPUs; the equivalent CPU cost is taken from \citet{Euclid_skies}.}                                                                                \\ \hline
            \begin{tabular}[c]{@{}l@{}}HalfDome\\ \citep{HalfDome}\end{tabular}          & \texttt{FastPM}    & $6\,144^3$                           & $3\,750$                             & $1.95\times10^{10}$                            & 2.63                                                                                       \\ \hline
            \begin{tabular}[c]{@{}l@{}}Gepetto\\ \citep{Euclid_skies}\end{tabular}       & \texttt{Pinocchio} & $2\,160^3$                           & $1\,200$                             & $1.52\times10^{10}$                            & 0.020                                                                                      \\ \hline
            \multirow{3}{*}{\textbf{This work}}                                          & \texttt{Gadget4}  & \multirow{2}{*}{$\bm{3\,840^3}$} & \multirow{2}{*}{$\bm{3\,840}$}   & \multirow{3}{*}{$\bm{8.62\times10^{10}}$}  & 59.68                                                                                      \\ \cline{2-2} \cline{6-6} 
                                                                                         & \textbf{sCOLA}     &                                      &                                      &                                            & \textbf{0.022}                                                                             \\ \cline{2-4} \cline{6-6} 
                                                                                         & tCOLA              & $1\,920^3$                           & $1\,920$                             &                                            & 0.029                                                                                      \\ \hline \hline
            \end{tabular}
            \caption{Computational cost of our simulations compared with that of other large cosmological simulations. The costs of our runs are given for lightcone simulations with 10 time steps and, for sCOLA, an oversimulation factor of 3.38 set by the buffer size. sCOLA reduces the cost by a factor of 120 to 2\,600 with respect to the other $N$-body codes, approaching the cost of \texttt{Pinocchio} while remaining an $N$-body solver. These ratios are indicative only: the runs compared differ in mass resolution, gravity solver, and hardware.
            }
            \label{tab:cpu}
        \end{table*}

The \texttt{Gadget4} \textit{Main} and \textit{Mini} simulations cost $3\,379\,000$ CPUh and $128\,200$ CPUh respectively. The monolithic tCOLA \textit{Mini} simulation cost 205 CPUh, which corresponds to an average of 9.5 CPU seconds per million particles per force step (hereafter CPU sec / M part / step). Overall, on the \textit{Mini} volume, tCOLA reduces the computational cost by a factor of 625 with respect to \texttt{Gadget4}. This reduction is due to (i) the reduced number of time steps allowed by COLA and (ii) the use of a PM force model.

For the \textit{Main} simulation up to $a_0=1$, an sCOLA box ran on average for 11.2 s (43.5 CPU seconds), corresponding to a total computational cost of 2\,690 CPUh.
This amounts to 15.6 CPU sec / M part / step when only the particles kept, i.e.\ those of the tile, are counted (CPU sec / M part kept / step), above the monolithic tCOLA value because of the oversimulation factor. When all particles evolved are counted, i.e.\ including the buffer, the cost is approximately 4.6 CPU sec / M part / step, below the tCOLA value because of the imperfect CPU efficiency of the monolithic run on large grids. The memory footprint of sCOLA was 170 MB / M part.
        
The cost per million particles evolved per time step is expected to be nearly independent of the sCOLA configuration; taking it to be the ${\sim}4.6$ CPU sec / M part / step measured for all of the configurations studied here, the total cost of an sCOLA simulation is proportional to the oversimulation factor (here 3.38). In this work, we chose the smallest tiles that achieve an accuracy of $1\%$ on the power spectrum, $L^\mathrm{tile}=64\,h^{-1}\,\mathrm{Mpc}$, to maximise the number of independent tiles, and hence the degree of parallelism.
Larger tiles, for instance $L^\mathrm{tile}=192\,h^{-1}\,\mathrm{Mpc}$ at fixed $L^\mathrm{buffer}=16\,h^{-1}\,\mathrm{Mpc}$, would have given an oversimulation factor of 1.59 and divided the total budget by $2.1$ (see Appendix~\ref{app:sCOLA_config}).

Because each sCOLA box is stopped as soon as the lightcone has left it, the boxes of the lightcone run ran for 5.4 s on average (20.8 CPU seconds), for a total of $1\,300$ CPUh: a 52\% reduction with respect to the run to $a_0=1$. This reduction is consistent with the predicted 57\% reduction in the number of force steps (Figs.~\ref{fig:lightcone_scola} and \ref{fig:fraction_tiles}). The difference is explained by the overhead due to I/O, which does not scale with the number of steps.
Overall, the computational cost of the sCOLA lightcone run is lower than that of our \texttt{Gadget4} run by a factor of $2\,600$. Four effects drive this reduction: (i) the reduced number of time steps allowed by COLA, (ii) the use of a PM force model, (iii) the absence of global communications between nodes and (iv) the restriction of the evolution to the observed spacetime region. It is partly offset by the oversimulation factor, which multiplies the number of particles evolved by 3.38.
These contributions can be separated using the runs of Section~\ref{sec:Simulations}. On the \textit{Mini} volume, effects (i), (ii) and (iii) together reduce the cost by a factor of 625 with respect to \texttt{Gadget4}. Restricting the evolution to the observed spacetime region adds a factor $2\,690/1\,300 = 2.1$. The tiling itself costs a net factor of 1.6, the oversimulation factor of 3.38 being partly compensated by the better CPU efficiency of the small sCOLA boxes (4.6 against 9.5 CPU sec / M part / step). The remaining factor of 3.3 is the higher per-particle cost of \texttt{Gadget4} on the \textit{Main} volume than on the \textit{Mini} (59.7 against 18.1 CPUh per million particles), and is therefore a property of the reference code rather than of sCOLA. Together, $625 \times 2.1 \times 3.3 / 1.6 \simeq 2\,600$.

Table~\ref{tab:cpu} compares the computational cost of our simulations with that of other large cosmological runs. sCOLA reaches a cost per particle comparable to that of approximate methods such as \texttt{Pinocchio} ($0.022$ versus $0.020$ CPUh per million particles) while retaining full $N$-body dynamics.
        
\section{Discussion}
\label{sec:Discussion}

We first ask what controlling the two sCOLA approximations has bought (Section~\ref{ssec:Accuracy improvements with respect to previous work}), and how a tiling should be chosen in practice (Section~\ref{ssec:Computational strategy and scalability}). We then consider what restricting the evolved volume to the observed spacetime region makes possible (Section~\ref{ssec:Lightcone simulations and applications}).

    \subsection{Accuracy improvements with respect to previous work}
    \label{ssec:Accuracy improvements with respect to previous work}

Compared to \citet{Leclercq2020sCOLA}, the present work improves the accuracy of sCOLA by better controlling its two approximations: the LPT buffer accounts for the inflow of particles, and the second-order gravitational potential improves the boundary conditions. In previous work, the sCOLA matter power spectrum differed from that of tCOLA at large scales (Fig.~6 of \citealt{Leclercq2020sCOLA}); this difference disappears with the LPT buffer and the second-order gravitational potential (Fig.~\ref{fig:sCOLA_approx}). The gain also allows smaller tiles and buffers: whereas \citet{Leclercq2020sCOLA} needed an sCOLA configuration of $L^\mathrm{tile}=125\,h^{-1}\,\mathrm{Mpc}$ and $L^\mathrm{buffer}=48.8\,h^{-1}\,\mathrm{Mpc}$ to reach $\sim 1\%$ agreement with tCOLA on the matter power spectrum, a configuration with $L^\mathrm{tile}=64\,h^{-1}\,\mathrm{Mpc}$ and $L^\mathrm{buffer}=16\,h^{-1}\,\mathrm{Mpc}$ now reaches $1\%$ agreement with \texttt{Gadget4}.

Accuracy improvements are directly visible in the density and velocity fields. In previous work, the density and velocity differences between sCOLA and tCOLA were of the order of $\pm 0.4$ in $\log_{10}(2+\delta)$ and $\pm 100\,\mathrm{km}\,\mathrm{s}^{-1}$, respectively (Figs.~3 and 4 of \citealt{Leclercq2020sCOLA}); here, between sCOLA and \texttt{Gadget4}, they are of the order of $\pm 0.15$ in $\log_{10}(2+\delta)$ and $\pm 50\,\mathrm{km}\,\mathrm{s}^{-1}$ (see the colour bar scales of Figs.~\ref{fig:density_lightcone} and \ref{fig:velocity_lightcone}). Although the tiling artefacts of \citet{Leclercq2020sCOLA} are no longer visible by eye, these residual differences are still preferentially located close to the tile edges and in high-density regions.
As reported in Section~\ref{ssec:Density and velocity fields}, these tile-related defects of the velocity field disappear when the buffer is enlarged; Appendix~\ref{app:sCOLA_config} shows more generally how the agreement with tCOLA improves with buffer size.

The velocity field also shows low-amplitude, spatially extended differences, which we interpret as a misestimation of bulk flows (of the order of $\pm 10\,\mathrm{km}\,\mathrm{s}^{-1}$). Both the density and velocity differences decrease with distance from the observer, as structures at higher redshift are less non-linear.

        The residual differences between sCOLA and \texttt{Gadget4} can be attributed to three approximations:
        \begin{enumerate}
            \item \textbf{The sCOLA tiling} leads to a bump at the scale of the sCOLA box in the density power spectrum, the velocity divergence power spectrum, and the density bispectrum. Its amplitude is controlled by the tile and buffer sizes.
			\item \textbf{The reduced number of time steps in COLA} leads to differences at wavenumbers $0.1 \lesssim k \lesssim 1\,h\,\mathrm{Mpc}^{-1}$ in the density power spectrum, the velocity divergence power spectrum, and the density bispectrum. These differences can be mitigated by increasing the number of time steps.
			\item \textbf{The PM force model} is responsible for the divergence at wavenumbers $k \gtrsim 1\,h\,\mathrm{Mpc}^{-1}$. Removing it requires a more accurate force model.
        \end{enumerate}

    \subsection{Computational strategy and scalability}
    \label{ssec:Computational strategy and scalability}

A key quantity in sCOLA is the oversimulation factor of equation~\eqref{eq:oversimulation}, which quantifies the extra volume simulated because of the buffers. The computational cost is roughly proportional to this ratio, which must therefore be balanced against accuracy requirements. Given the available RAM, the target mass resolution, and the accuracy requirements, one can use the Pareto fronts of Figs.~\ref{fig:pareto_oversimulation} and \ref{fig:pareto_ram} to find the $L^\mathrm{tile}$ and $L^\mathrm{buffer}$ that minimise the oversimulation factor. For instance, with 2~GB of RAM per sCOLA box, the maximum number of particles per box is $230^3$; at the mass resolution of this work ($8.62\times10^{10}\,h^{-1}\,\mathrm{M}_\odot$) and for a required accuracy of $1\%$, the configuration that minimises the oversimulation factor is then $L^\mathrm{tile}=192\,h^{-1}\,\mathrm{Mpc}$ and $L^\mathrm{buffer}=16\,h^{-1}\,\mathrm{Mpc}$, with an oversimulation factor of 1.59. With this configuration, the \textit{Main} lightcone simulation would then cost an estimated 600~CPUh.

The tiling strategy enables asynchronous and perfectly parallel computations (see \citealt{Leclercq2020sCOLA} for a detailed discussion). Unlike traditional large $N$-body simulations, this approach does not require tightly synchronised high-performance computing resources. Instead, it allows a flexible use of distributed computing resources and is resilient to node failure. On the other hand, it does not exploit the fast inter-node communication capabilities typical of large supercomputers. The memory footprint of a single sCOLA box is also small, 150~MB for the configuration used here, so that sCOLA could in principle run on a wide range of hardware, and even be distributed over volunteer computing resources.

    \subsection{Lightcone simulations and applications}
    \label{ssec:Lightcone simulations and applications}

Cosmological surveys measure the properties of galaxies over a fraction of the sky and within a range of redshifts. Their footprints do not fill their bounding cube (the smallest cube containing the survey geometry). For instance, the Galactic north cap covered by \textit{Euclid} spectroscopic data, defined as a truncated spherical cone sector covering the redshift range $0.84\leq z\leq 1.88$ and a Galactic latitude $b > 23^\circ$ \citep{Euclid_Wide_Survey}, has a volume of $42.9\,(h^{-1}\,\mathrm{Gpc})^3$, only $29\%$ of the volume of its bounding cube (Appendix~\ref{app:northcap}).

The observed spacetime region is therefore further restricted by the survey geometry: of the past lightcone contained in the simulation box, only the part covered by the footprint is of interest. The sCOLA tiling makes it possible to evolve only the tiles that overlap that region, so that the cost of a simulation is reduced accordingly. Evolving only the observed spacetime region thus combines a restriction in time, set by the lightcone, and a restriction in space, set by the survey geometry. This spatial restriction also removes the need to replicate the box: since only the tiles overlapping the footprint are evolved, the bounding cube can be made as large as the survey requires while the evolution is paid for over the observed volume alone. The initial conditions are the exception: the LPT potentials must still be computed over the entire cube (Section~\ref{ssec:Review of sCOLA}), so the memory footprint of that step is set by the bounding volume and not by the survey.

With an allocation of $\sim 10^{7}$~CPUh, comparable to that of state-of-the-art projects \citep{EuclidMock,Euclid_skies,HalfDome}, one could run $\sim 17\,000$ \textit{Main} lightcone simulations in the $L^\mathrm{tile}=192\,h^{-1}\,\mathrm{Mpc}$, $L^\mathrm{buffer}=16\,h^{-1}\,\mathrm{Mpc}$ configuration. Extrapolating to the \textit{Euclid} north cap volume, the same budget would allow for $\sim 25\,000$ north cap simulations (Appendix~\ref{app:northcap}) at the mass resolution of this work. On the other hand, improving the mass resolution to resolve the smallest halos containing tracers detectable by the survey would require many more particles (as in \citealt{EuclidMock}) and therefore a much higher cost per simulation. 

Combined with a prescription for populating the dark matter field with halos and galaxies, sCOLA therefore opens the possibility of producing large ensembles of mock catalogues. Such ensembles are required to estimate the covariance matrices of two-point and higher-order statistics \citep{Hartlap2007,Patchy_mocks,DESI_covariance}, to perform simulation-based inference (SBI, also known as implicit-likelihood inference, ILI) \citep{SimBig,SBI_howmanysims}, to test modified gravity models \citep{MG_COLA_mocks,MGGLAM}, and to probe relativistic effects such as gravitational redshift \citep{GravRedshift_clusters,Relativistic_BOSS,Relativistic_sims}. All these applications need mocks that are deep as well as numerous: a box shallower than the survey suffers from super-sample covariance \citep{Supersample_cov} and, once replicated to fill the volume, from spurious correlations between its copies \citep{BoxReplicationWL}.

\section{Conclusion}
\label{sec:Conclusion}

    Lightcone simulations at survey scale have so far been affordable one at a time. We have shown that sCOLA lifts that constraint: because the volume is tiled into independent boxes, each box can be stopped as soon as the observer's past lightcone has left it, and only the boxes intersecting the region of interest need be evolved at all, so that evolving only the observed spacetime region becomes a property of the tiling rather than a scheme added on top of it. We also improved the tiling itself, with an LPT buffer that approximates the inflow of particles into a box and a second-order approximation to the gravitational potential at its boundaries; together they remove the large-scale inaccuracies observed by \citet{Leclercq2020sCOLA} and allow substantially smaller buffers and tiles, which is what makes the method cheap as well as perfectly parallel.

    With respect to \texttt{Gadget4}, sCOLA reaches percent-level accuracy in the matter power spectrum up to $k \simeq 0.4\,h\,\mathrm{Mpc}^{-1}$, $2\%$ agreement in the equilateral bispectrum up to the same scale, and $1.5\%$ agreement in the velocity divergence power spectrum up to $k \simeq 0.6\,h\,\mathrm{Mpc}^{-1}$, at a computational cost reduced by a factor of $2\,600$. Four effects drive this reduction:
    \begin{enumerate}[leftmargin=0.4cm, itemsep=1mm]
        \item \textbf{Changing the frame of reference}: sCOLA is an extension of COLA, in which we solve the equations of motion of particles in the reference frame of their 2LPT trajectories; fewer time steps are then needed to reach a given accuracy than in a standard $N$-body integration \citep{Tassev2015}.
        \item \textbf{Using a particle-mesh force model}: gravitational forces are evaluated on a particle-mesh grid rather than with a tree, which makes each time step cheaper, at the price of accuracy below the scales resolved by the mesh.
        \item \textbf{Avoiding global communications}: thanks to appropriate approximations at the tile boundaries (a buffer of simulated particles, an LPT buffer, and Dirichlet boundary conditions on the gravitational potential), each tile can be simulated independently of all the others, with no communication at all: the algorithm is perfectly parallel \citep{Leclercq2020sCOLA}.
        \item \textbf{Evolving only the observed spacetime region}: we simulate only the tiles that intersect the region of interest, and evolve each of them only until the lightcone has left it (the contribution of the present paper).
    \end{enumerate}

    The perfect parallelism and low cost of sCOLA make it particularly well suited to generating the large ensembles of simulations required by modern cosmological analyses: for instance, we estimate that $\sim 25\,000$ lightcone simulations of the \textit{Euclid} Galactic north cap at a mass resolution of $8.62\times10^{10}\,h^{-1}\,\mathrm{M}_\odot$ would require a budget of $\sim 10^{7}$~CPUh. Such ensembles are essential for covariance estimation, simulation-based inference, and tests of fundamental physics.

    Three directions follow naturally. The halo and galaxy statistics of sCOLA lightcones, in redshift space, remain to be validated against a full $N$-body reference, since it is halos, not the matter field, that mock catalogues are built from; there, ten time steps may not suffice, since even a 20-step COLA integrator, which describes the density field around the most massive clusters accurately, does not yield converged virial masses for them \citep{Stopyra2024}. The accuracy limit set by the particle-mesh force model, visible at $k \gtrsim 1\,h\,\mathrm{Mpc}^{-1}$, could be pushed to smaller scales by a more accurate short-range gravitational solver. Finally, the present framework could be combined with an emulated frame of reference to further reduce its computational cost without sacrificing asymptotic convergence to the physical solution \citep[see][]{Bartlett2025tCOCA}. With these in hand, producing the ensembles of galaxy mock catalogues that Stage-IV surveys call for becomes a question of allocation rather than of method.

\appendix

\section{LPT buffer}\label{app:LPT_buffer}

\citet{Leclercq2020sCOLA} quantified the error caused by the sCOLA approximations on the density field $\delta \approx \delta_\mathrm{int}$ and on the gravitational boundary conditions $\Phi_\mathrm{BCs} \approx \Phi_\mathrm{LEP}$ (their Fig.~7). The dominant source of error was the approximation on the density field $\delta \approx \delta_\mathrm{int}$. The discrepancy between $\delta_\mathrm{int}$ and $\delta$ is mostly due to the lack of particle inflow while the Lagrangian sCOLA box deforms (see Fig.~\ref{fig:sCOLA_concept}). This flow imbalance leads to defects in the gravitational potential near the boundaries of the box, which propagate as the simulation progresses. To address this issue, we implemented an additional buffer of particles around the sCOLA box, the LPT buffer, to approximate the missing inflow. 

    Particles in this LPT buffer follow the 2LPT trajectories of equation~\eqref{eq:2LPT_pos}; no forces are computed for them. We compute the positions of the LPT-buffer particles on the fly, and project their mass onto the PM density grid using the same density assignment scheme as for particles of the sCOLA box (CiC here). Formally, using the notation of \citet{Leclercq2020sCOLA}, where $\mathrm{B}^\mathrm{sCOLA}$ denotes the density assignment operator, $\left\lbrace \bm{x}(a) \right\rbrace_\mathrm{sCOLA}$ the set of particles of the sCOLA box, and $\left\lbrace \bm{x}(a) \right\rbrace_\mathrm{LPT~buffer}$ that of the LPT buffer, we replaced the prescription \citep[][equation (19)]{Leclercq2020sCOLA}
\begin{equation}
\delta_\mathrm{int}(a) \equiv \mathrm{B}^\mathrm{sCOLA}\left(\left\lbrace \bm{x}(a) \right\rbrace_\mathrm{sCOLA} \right)
\end{equation}
by
\begin{align}
\delta_\mathrm{int}(a) & \equiv \mathrm{B}^\mathrm{sCOLA}\left(\left\lbrace \bm{x}(a) \right\rbrace_\mathrm{sCOLA} \cup \left\lbrace \bm{x}(a) \right\rbrace_\mathrm{LPT~buffer} \right).
\end{align}
    Figure~\ref{fig:sCOLA_concept} shows the overall structure of the new tiling scheme: each sCOLA box is composed of a tile and a buffer of particles (with PM trajectory), surrounded by an LPT buffer.

The LPT buffer must be wide enough to contain the particles that can enter the sCOLA box in the course of the run: its width should therefore exceed the distance a particle travels between the initial time and $a_0=1$. That distance can be estimated within perturbation theory: at leading order the displacement is $-D_1(a)\,\boldsymbol{\varPsi}_1$, so that the variance of one of its Cartesian components at $a_0=1$ is
\begin{equation}
    \sigma_\varPsi^2 = \frac{1}{6\pi^2}\int P_\mathrm{L}(k)\,\drm k ,
    \label{eq:rms_displacement}
\end{equation}
where $P_\mathrm{L}$ is the linear matter power spectrum at $a_0=1$; the second-order term contributes only a small correction to $\sigma_\varPsi$ \citep{Bouchet1995}. For the cosmology of Section~\ref{ssec:General setup}, equation~\eqref{eq:rms_displacement} gives $\sigma_\varPsi \simeq 5.9\,h^{-1}\,\mathrm{Mpc}$ per Cartesian component. The LPT buffer adopted in this work, $L^\mathrm{LPT~buffer}=8\,h^{-1}\,\mathrm{Mpc}$, that is $N_\mathrm{p}^\mathrm{LPT~buffer}=8$ layers of particles (Section~\ref{ssec:sCOLA}), therefore covers about $1.4\,\sigma_\varPsi$, and so accounts for the inflow of the large majority of the particles that cross the boundary of the sCOLA box. The residual effect of this approximation on the power spectrum is quantified in Appendix~\ref{app:improvements}.

\section{Second-order gravitational potential approximation term $\Phi_2$}\label{app:Phi2}

We start the derivation from the Vlasov--Poisson system in the single-stream regime
\begin{equation}
\Delta_{\bm{x}} \Phi(\bm{x},\tau) = \delta(\bm{x},\tau),
\label{eq:Poisson-single-stream}
\end{equation}
where $\Phi$ absorbs the usual normalisation $4\pi \G a^2(\tau) \bar{\rho}(\tau)$.

    We consider the case of an Einstein–de Sitter universe. In second-order Eulerian (standard) perturbation theory, the density contrast is $\delta(\bm{k},\tau) = D_1(\tau) \delta_1(\bm{k}) + D_1^2(\tau) \delta_2(\bm{k})$, with
    \begin{equation}
        \delta_2(\bm{k}) = \int F_2(\bm{k}',\bm{k}-\bm{k}') \delta_1(\bm{k}') \delta_1(\bm{k}-\bm{k}') \, \drm^3 \bm{k}',
        \label{eq:delta_2EPT}
    \end{equation}
    where
    \begin{equation}
        F_2(\bm{k}_1, \bm{k}_2) \equiv \frac{5}{7} + \frac{1}{2} \left( \frac{k_1}{k_2} + \frac{k_2}{k_1} \right) \mu + \frac{2}{7} \mu^2,  \quad \mu \equiv \frac{\bm{k}_1 \cdot \bm{k}_2}{k_1 k_2}.
    \end{equation}

    We aim at building Eulerian perturbation theory at second order in the gravitational potential, such that
    \begin{equation}
        \Phi(\bm{k},\tau) = D_1(\tau) \Phi_1(\bm{k}) + D_1^2(\tau) \Phi_2(\bm{k}) ,
    \end{equation}
    with $\Phi_1$ rather than $\delta_1$ as the fundamental field. The first term corresponds to the linearly-evolving potential (LEP) approximation. The potential $\Phi_1$ and the linear density contrast $\delta_1$ are linked by the Poisson equation:
    \begin{equation}
        \Delta_{\bm{x}} \Phi_1(\bm{x}) = \delta_1(\bm{x}), \quad \delta_1(\bm{k}) = -k^2 \Phi_1(\bm{k}),
        \label{eq:Poisson-linear}
    \end{equation}
    therefore $\Phi_1$ coincides with $\phi^{(1)}$, the gravitational potential that appears in first-order Lagrangian perturbation theory and obeys (in Lagrangian coordinates) $\Delta_{\bm{q}} \phi^{(1)}(\bm{q}) = \delta_1(\bm{q})$. Furthermore, equations~\eqref{eq:Poisson-single-stream} and \eqref{eq:Poisson-linear} require
    \begin{equation}
        \Phi_2(\bm{k}) = -\frac{1}{k^2} \delta_2(\bm{k}).
        \label{eq:Phi2_delta2}
    \end{equation}
    From equations~\eqref{eq:delta_2EPT} and \eqref{eq:Phi2_delta2}, we obtain the formal expression of $\Phi_2(\bm{k})$ as a convolution of $\Phi_1(\bm{k})$:
    \begin{equation}
        \Phi_2(\bm{k}) = \int K_2(\bm{k}',\bm{k}-\bm{k}') \Phi_1(\bm{k}') \Phi_1(\bm{k}-\bm{k}') \, \drm^3 \bm{k}',
    \end{equation}
    where the kernel for the gravitational potential $K_2$ is related to the kernel for the density contrast $F_2$ by:
    \begin{equation}
        K_2(\bm{k}_1, \bm{k}_2) \equiv -F_2(\bm{k}_1, \bm{k}_2) \frac{k_1^2 k_2^2}{\Vert\bm{k}_1+\bm{k}_2\Vert^2} =-F_2(\bm{k}_1, \bm{k}_2) \frac{k_1^2 k_2^2}{k^2},
    \end{equation}
    with $k^2 \equiv \Vert\bm{k}_1+\bm{k}_2\Vert^2$. In an Einstein–de Sitter universe, $K_2$ is a purely spatial kernel (as is $F_2$).

    We now express $K_2$ in terms of the kernels that correspond to the operators $\phi^{(2)}$ and $\Vert\boldsymbol{\varPsi}_1\Vert^2$, which are two quadratic invariants of $\phi^{(1)}$ that appear in LPT:
    \begin{equation}
        \boldsymbol{\varPsi}_1(\bm{q}) = \boldsymbol{\nabla}_{\bm{q}} \phi^{(1)},
        \label{eq:1LPTdisplacement}
    \end{equation}
    \begin{align}
        \Delta_{\bm{q}} \phi^{(2)}(\bm{q}) & = \sum_{\ell>m} \left[ \phi^{(1)}_{,\ell\ell}\phi^{(1)}_{,mm} - \left(\phi^{(1)}_{,\ell m}\right)^2 \right](\bm{q}) \nonumber\\ 
        & = \frac{1}{2} \left[ \left(\Delta_{\bm{q}} \phi^{(1)}\right)^2 - \left(\partial_\ell \partial_m \phi^{(1)}\right)^2 \right](\bm{q}). \label{eq:LPTpotential2}
    \end{align}
    In Fourier space, in Lagrangian coordinates, we have
    \begin{align}
        \left[ \Delta_{\bm{q}} \phi^{(1)} \right](\bm{k}) & = -k^2 \phi^{(1)}(\bm{k}), \\
        \left[\partial_\ell \partial_m \phi^{(1)} \right](\bm{k}) & = -k_\ell k_m \phi^{(1)}(\bm{k}),
    \end{align}
    so the kernel for $\Delta_{\bm{q}} \phi^{(2)}$ is
    \begin{equation}
        K_{\Delta \phi^{(2)}}(\bm{k}_1, \bm{k}_2) \equiv \frac{1}{2} \left[ k_1^2 k_2^2 - \left( \bm{k}_1 \cdot \bm{k}_2 \right)^2 \right] = \frac{k_1^2 k_2^2}{2} \left( 1 - \mu^2 \right).
    \end{equation}
    Since $\left[\Delta \phi^{(2)} \right](\bm{k}) = -k^2 \phi^{(2)}(\bm{k})$, the kernel for $\phi^{(2)}$ itself is
    \begin{equation}
        K_{\phi^{(2)}}(\bm{k}_1, \bm{k}_2) \equiv -\frac{k_1^2 k_2^2}{2 k^2} \left( 1 - \mu^2 \right).
    \end{equation}
    Furthermore, $\left(\boldsymbol{\varPsi}_1\right)_\ell (\bm{k}) = \irm k_\ell \phi^{(1)}(\bm{k})$, so the kernel for $\Vert\boldsymbol{\varPsi}_1\Vert^2 = \boldsymbol{\varPsi}_1 \cdot \boldsymbol{\varPsi}_1$ is
    \begin{equation}
        K_{\Vert\boldsymbol{\varPsi}_1\Vert^2}(\bm{k}_1, \bm{k}_2) = \sum_\ell (\irm k_{1\, \ell})(\irm k_{2\, \ell}) = -\bm{k}_1 \cdot \bm{k}_2 = - k_1 k_2 \mu.
    \end{equation}

    Let us seek constants $A$ and $B$ such that
    \begin{equation}
        K_2(\bm{k}_1, \bm{k}_2) = A K_{\phi^{(2)}}(\bm{k}_1, \bm{k}_2) + B K_{\Vert\boldsymbol{\varPsi}_1\Vert^2}(\bm{k}_1, \bm{k}_2).
    \end{equation}
    This equation is successively equivalent to:
    \begin{equation}
        \frac{k^2}{k_1^2 k_2^2} \left[ A K_{\phi^{(2)}} + B K_{\Vert\boldsymbol{\varPsi}_1\Vert^2} \right] (\bm{k}_1, \bm{k}_2) = -F_2(\bm{k}_1, \bm{k}_2),
    \end{equation}
    \begin{equation}
        -\frac{A}{2}\left( 1-\mu^2 \right) - B \mu \frac{k^2}{k_1 k_2} = -\frac{5}{7} - \frac{1}{2} \left( \frac{k_1}{k_2} + \frac{k_2}{k_1} \right) \mu - \frac{2}{7} \mu^2,
    \end{equation}
    \begin{align}
        -\frac{A}{2} - B \left( \frac{k_1}{k_2} + \frac{k_2}{k_1} \right) \mu & + \left( \frac{A}{2}-2B \right) \mu^2 \nonumber \\
        & = -\frac{5}{7} - \frac{1}{2} \left( \frac{k_1}{k_2} + \frac{k_2}{k_1} \right) \mu - \frac{2}{7} \mu^2,
    \end{align}
    using $\frac{k^2}{k_1 k_2} = \frac{k_1}{k_2} + \frac{k_2}{k_1} + 2 \mu$.
    Matching the independent terms in $1$, $\mu$ and $\mu^2$ gives
    \begin{equation}
        -\frac{A}{2} = -\frac{5}{7}, \quad -B = -\frac{1}{2}, \quad \frac{A}{2} - 2B = -\frac{2}{7},
    \end{equation}
    which are mutually consistent and yield
    \begin{equation}
        A = \frac{10}{7}, \quad B=\frac{1}{2}.
    \end{equation}
    The final result is
    \begin{equation}
        K_2(\bm{k}_1, \bm{k}_2) = \frac{10}{7} K_{\phi^{(2)}}(\bm{k}_1, \bm{k}_2) + \frac{1}{2} K_{\Vert\boldsymbol{\varPsi}_1\Vert^2}(\bm{k}_1, \bm{k}_2),
    \end{equation}
    which corresponds in configuration space to the following expression for the gravitational potential at second order:
    \begin{equation}
        \Phi_2(\bm{q}) = \frac{10}{7} \phi^{(2)}(\bm{q}) + \frac{1}{2}\Vert\boldsymbol{\varPsi}_1(\bm{q})\Vert^2.
    \end{equation}

    This expression is valid in Lagrangian coordinates. The boundary conditions are, however, required at the Eulerian positions $\bm{x}$ of the faces of the sCOLA box, whereas $\phi^{(2)}$ and $\boldsymbol{\varPsi}_1$ are known on the Lagrangian grid. Using the expression above with $\bm{x}$ in place of $\bm{q}$ amounts to neglecting the Jacobian of the transformation between the two sets of coordinates, which only affects the result at third order. Indeed, the displacement $\bm{x}-\bm{q}$ is of first order at leading order (equation~\eqref{eq:2LPT_pos}); therefore, any field $F$ that is already quadratic in $\phi^{(1)}$ satisfies
\begin{equation}
F(\bm{x}) = F(\bm{q}) + (\bm{x}-\bm{q})\cdot\boldsymbol{\nabla}_{\bm{q}} F(\bm{q}) + \mathcal{O}\!\left(\Vert\bm{x}-\bm{q}\Vert^2\right) ,
\end{equation}
whose first correction is of third order. The same counting holds for the derivative operators, which satisfy $\partial/\partial x_i = \partial/\partial q_i + \mathcal{O}(\bm{x}-\bm{q})$ and thus modify the equations defining $\phi^{(1)}$ and $\phi^{(2)}$ one order beyond the order at which they are used. Since third-order terms are in any case discarded in the second-order expansion of $\Phi$, keeping track of the Jacobian would not improve the accuracy of the approximation.

\section{Accuracy gains from the LPT buffer and the second-order boundary conditions}\label{app:improvements}

    \begin{figure}
        \centering
        \includegraphics[width=1\linewidth]{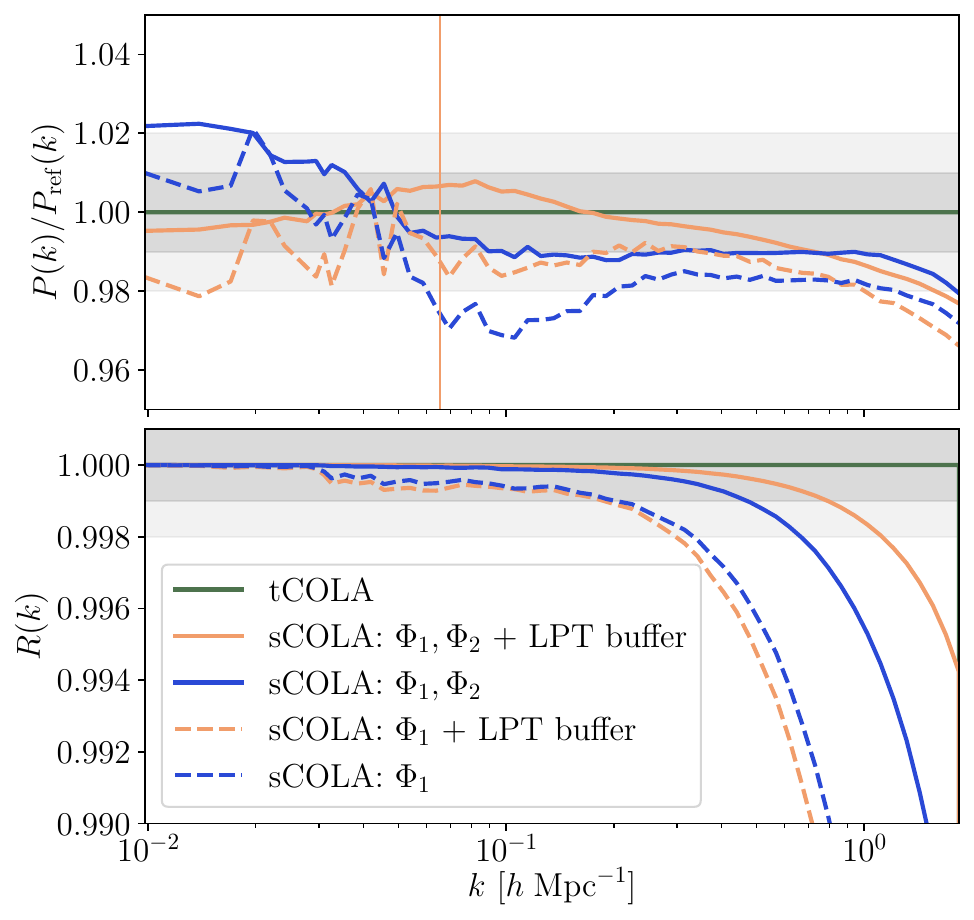}
        \caption{Effects of the LPT buffer and second-order gravitational potential term $\Phi_2$ on the sCOLA power spectrum and cross-correlations compared to monolithic tCOLA, for a configuration $L=640\,h^{-1}\,\mathrm{Mpc}$, $L^\mathrm{tile}=64\,h^{-1}\,\mathrm{Mpc}$, $L^\mathrm{buffer}=16\,h^{-1}\,\mathrm{Mpc}$, $L^\mathrm{LPT~buffer}=8\,h^{-1}\,\mathrm{Mpc}$, and the same mass resolution as the \textit{Main} simulation. The two improvements combined bring the power spectrum error down from 3\% to 1\%, remove the bias at large scales, and push the scale at which the cross-correlations diverge from $k\simeq 0.2$ to $0.8\,h\,\mathrm{Mpc}^{-1}$.
        The vertical line in the top panel marks the fundamental mode of the sCOLA box, $k_\mathrm{sCOLA}\equiv 2\pi/L^\mathrm{sCOLA}\simeq 0.065\,h\,\mathrm{Mpc}^{-1}$, with $L^\mathrm{sCOLA}=L^\mathrm{tile}+2L^\mathrm{buffer}=96\,h^{-1}\,\mathrm{Mpc}$.}
        \label{fig:sCOLA_approx}
    \end{figure}

    With the LPT buffer added, sCOLA has two approximations with respect to tCOLA: (i) the trajectories of inflow particles are 2LPT trajectories rather than tCOLA trajectories, and (ii) the boundary conditions on the gravitational potential are those of the approximation $\Phi_\mathrm{BCs}(a)$ rather than the exact $\Phi(a)$. We tested these two approximations separately, by replacing the trajectories of inflow particles by tCOLA trajectories, and the boundary conditions by the tCOLA potential. We found that once the LPT buffer is enabled, the dominant source of error is the boundary conditions on the potential; this is why we sought an approximation at second order.

    Figure~\ref{fig:sCOLA_approx} quantifies the gain brought by each improvement on its own and by the two combined, in a simulation of $N_\mathrm{p}^3 = 640^3$ particles in a box of $L=640\,h^{-1}\,\mathrm{Mpc}$. The sCOLA configuration is $L^\mathrm{tile}=64\,h^{-1}\,\mathrm{Mpc}$, $L^\mathrm{buffer}=16\,h^{-1}\,\mathrm{Mpc}$ and $L^\mathrm{LPT~buffer}=8\,h^{-1}\,\mathrm{Mpc}$. Without either of them, sCOLA is within $3\%$ of monolithic tCOLA on the power spectrum and the cross-correlations diverge at $k\simeq 0.2\,h\,\mathrm{Mpc}^{-1}$. The LPT buffer alone brings the power spectrum within $2\%$ but leaves the cross-correlations unchanged, which shows that the missing inflow is not the only source of error. The second-order gravitational potential alone also brings the power spectrum within $2\%$, and pushes the divergence of the cross-correlations to $k\simeq 0.5\,h\,\mathrm{Mpc}^{-1}$. With both, the sCOLA power spectrum converges to that of tCOLA at large scales and stays within $1\%$ of it, and the cross-correlations diverge at $k\simeq 0.8\,h\,\mathrm{Mpc}^{-1}$.

\section{sCOLA configurations: tile and buffer optimal choices}\label{app:sCOLA_config}

    The comparisons presented in this appendix are carried out on a dedicated test volume, which we refer to as the \textit{Micro} simulation: a box of side $L^\mathrm{Micro}=960\,h^{-1}\,\mathrm{Mpc}$ with $N_\mathrm{p}^\mathrm{Micro}=960$ particles and $N_\mathrm{pm}^\mathrm{Micro}=1\,920$ PM grid cells per dimension, that is, the same mass and force resolutions as the \textit{Main} and \textit{Mini} simulations of Section~\ref{sec:Simulations} ($d_\mathrm{p}\equiv L/N_\mathrm{p}=1\,h^{-1}\,\mathrm{Mpc}$, $m_\mathrm{p}=8.62\times10^{10}\,h^{-1}\,\mathrm{M}_\odot$). The cosmological parameters, the initial conditions, the starting scale factor $a_{\mathrm{i}}=0.05$ and the 10 KDK time steps are also those of Section~\ref{sec:Simulations}; the simulation is evolved to $a_0=1$ and no lightcone is built. This volume is small enough to be run monolithically, which provides the tCOLA reference $P_\mathrm{ref}$ used below, and its side is divisible by all the tile sizes considered here.

    To compare sCOLA configurations quantitatively, we define the distance of a configuration to the tCOLA reference as the sum of a distance on the power spectrum and one on the cross-correlations:
    \begin{align}\label{eq:distance_tCOLA}
        D_{\mathrm{ref}}\equiv &\sqrt{\frac{1}{N_k}\sum_{k_i<k_\mathrm{s}}\left[1-\frac{P(k_i)}{P_\mathrm{ref}(k_i)}\right]^2}\nonumber\\
        &+100\times \sqrt{\frac{1}{N_k}\sum_{k_i<k_\mathrm{s}}\left[1-R(k_i)\right]^2}
    \end{align}
    with $\left\lbrace k_i \right\rbrace$ the wavenumber bins, $k_\mathrm{s}\equiv 1.0\, h\,\mathrm{Mpc}^{-1}$ the maximum wavenumber considered, $N_k$ the number of modes below $k_\mathrm{s}$, $P$ and $P_\mathrm{ref}$ the sCOLA and tCOLA power spectra, and $R$ their cross-correlation.

    sCOLA has two more parameters than a monolithic tCOLA simulation: the size of the tile $L^\mathrm{tile}$ and the size of the buffer $L^{\mathrm{buffer}}$. We keep the particle resolution $d_\mathrm{p}\equiv\frac{L}{N_\mathrm{p}}$ (or equivalently the mass resolution) and the PM grid resolution $d_\mathrm{pm}\equiv\frac{L}{N_\mathrm{pm}}$ fixed. When exploring the ($L^\mathrm{tile}$, $L^\mathrm{buffer}$) space, four constraints must be satisfied:
    \begin{enumerate}[leftmargin=0.5cm, itemsep=-0.5mm]
        \item The number of tiles per axis $N_\mathrm{tiles}=\frac{L}{L^\mathrm{tile}}$ must be an integer;
        \item The number of particles per axis in the tile $N_\mathrm{p}^\mathrm{tile}=\frac{L^\mathrm{tile}}{d_\mathrm{p}}$ must be an integer;
        \item The number of particles per axis in the buffer $N_\mathrm{p}^\mathrm{buffer}=\frac{L^\mathrm{buffer}}{d_\mathrm{p}}$ must be an integer;
        \item The sCOLA box PM grid size $N_\mathrm{pm}^\mathrm{sCOLA}=\frac{L^\mathrm{tile}+2L^\mathrm{buffer}}{d_\mathrm{pm}}$ must have a simple prime factor decomposition for the DST operations to be fast. More precisely, the potential in a sCOLA box is obtained with type-I DSTs, whose Dirichlet boundary planes lie one cell outside the stored grid: the grid holds $N_\mathrm{pm}^\mathrm{sCOLA}-1$ cells and the logical size of the transform, which governs its cost, is $N_\mathrm{pm}^\mathrm{sCOLA}$. We therefore require this number to be $7$-smooth, $N_\mathrm{pm}^\mathrm{sCOLA}=2^{\alpha_2}\,3^{\alpha_3}\,5^{\alpha_5}\,7^{\alpha_7}$, and to be dominated by its factors of $2$, in the sense that $\alpha_3+2\alpha_5+3\alpha_7<\alpha_2$, each odd prime being weighted by the exponent of the power of $2$ closest to it ($3\simeq2$, $5\simeq2^2$, $7\simeq2^3$). When the requested grid does not satisfy this criterion, \texttt{Simbelmynë} substitutes for it the smallest admissible size at or above it, at fixed PM cell size $d_\mathrm{pm}$, so that the force resolution is left unchanged. The substitution is never downwards: an admissible size below $N_\mathrm{pm}^\mathrm{sCOLA}$, even when it is the closest one, would leave the PM grid too small to cover the sCOLA particle box, and its outermost particles would then receive no force and deposit only part of their mass.
    \end{enumerate}

\begin{figure}
        \centering
        \includegraphics[width=1\linewidth]{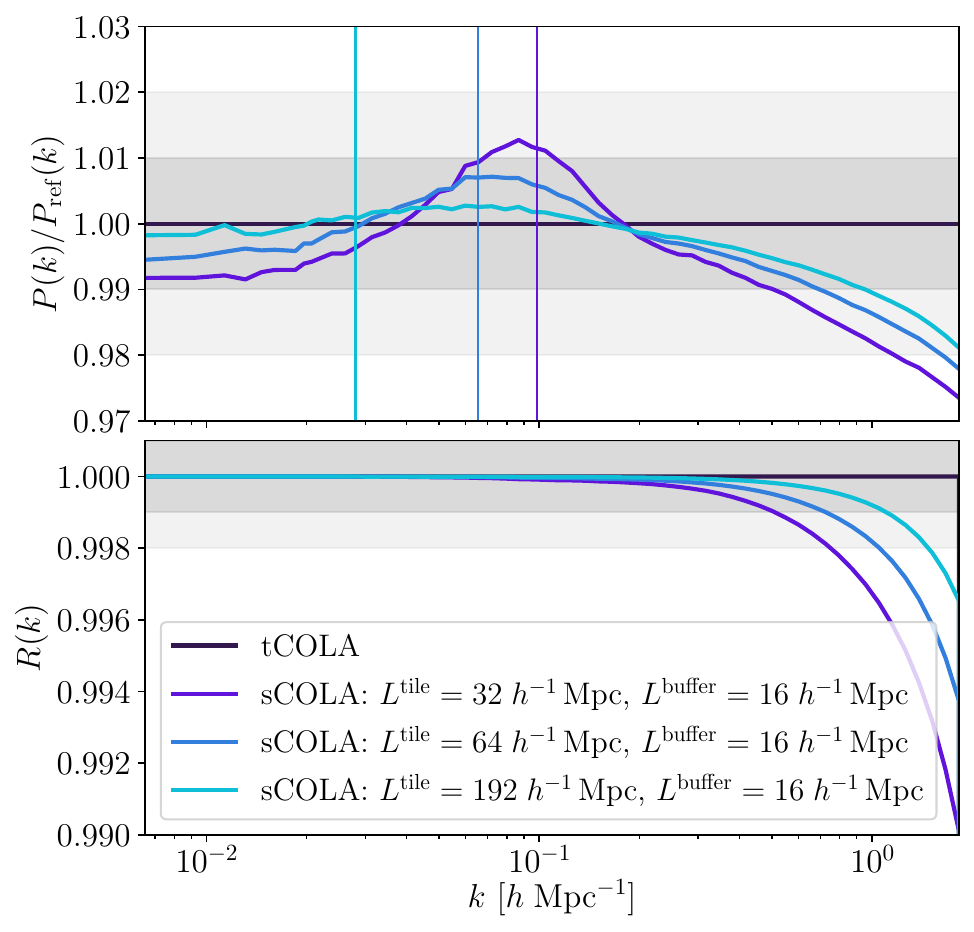}
        \caption{sCOLA power spectra and cross-correlations for the \textit{Micro} simulation at $a_0=1$ compared to tCOLA, for different tile sizes $L^\mathrm{tile}$ at a fixed buffer size $L^\mathrm{buffer}=16\,h^{-1}\,\mathrm{Mpc}$. Increasing the tile size decreases the distance to tCOLA: the sCOLA approximation introduces a bump in the power spectrum, which moves towards smaller $k$ and loses amplitude as $L^\mathrm{tile}$ increases. The bump sits close to $k_\mathrm{sCOLA}\equiv 2\pi/\left(L^\mathrm{tile}+2L^\mathrm{buffer}\right)$, the fundamental mode of the sCOLA box, shown as a vertical line for each configuration.}
        \label{fig:sCOLA_tile_size}
    \end{figure}

\begin{figure}
        \centering
        \includegraphics[width=1\linewidth]{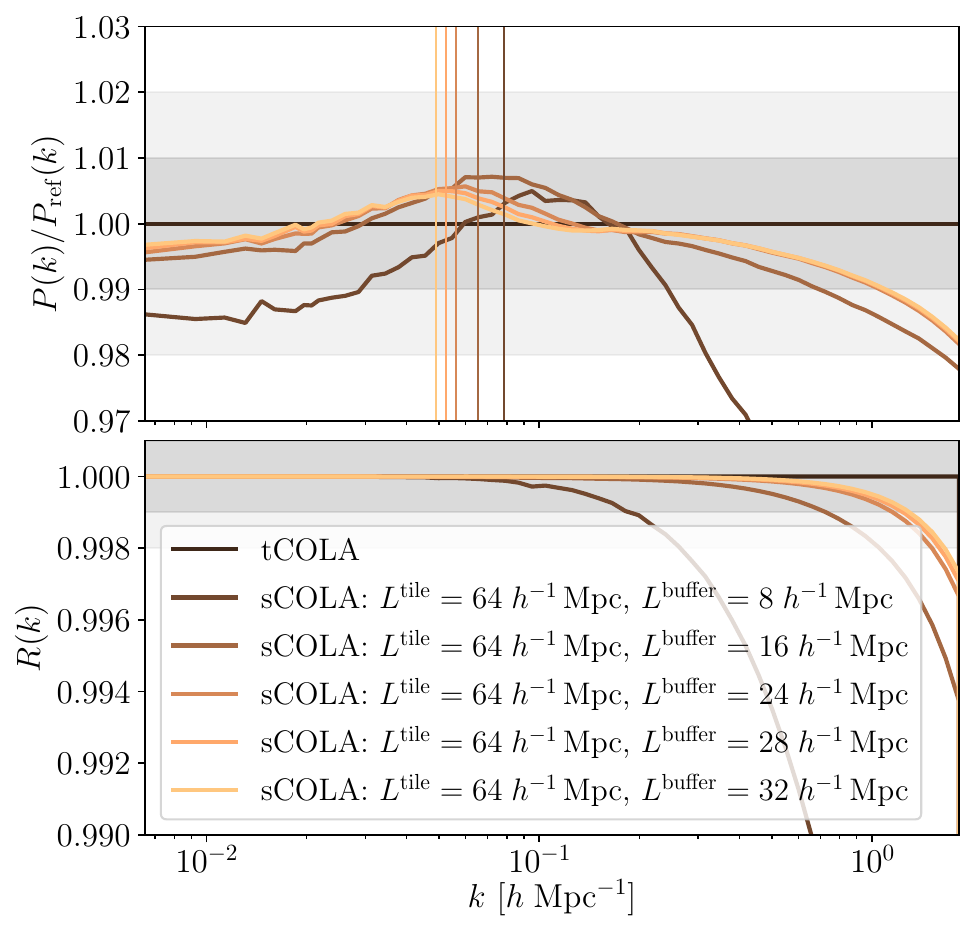}
        \caption{sCOLA power spectra and cross-correlations for the \textit{Micro} simulation at $a_0=1$ compared to tCOLA, for different buffer sizes $L^\mathrm{buffer}$ at a fixed tile size $L^\mathrm{tile}=64\,h^{-1}\,\mathrm{Mpc}$. Increasing the buffer size decreases the distance to tCOLA: as in Fig.~\ref{fig:sCOLA_tile_size}, the bump introduced by the sCOLA approximation moves towards smaller $k$ and loses amplitude as $L^\mathrm{buffer}$ increases, and sits close to $k_\mathrm{sCOLA}$ (vertical lines).}
        \label{fig:sCOLA_buffer_size}
    \end{figure}

\begin{figure}
        \vspace{0.5cm}
        \centering
        \includegraphics[width=1\linewidth]{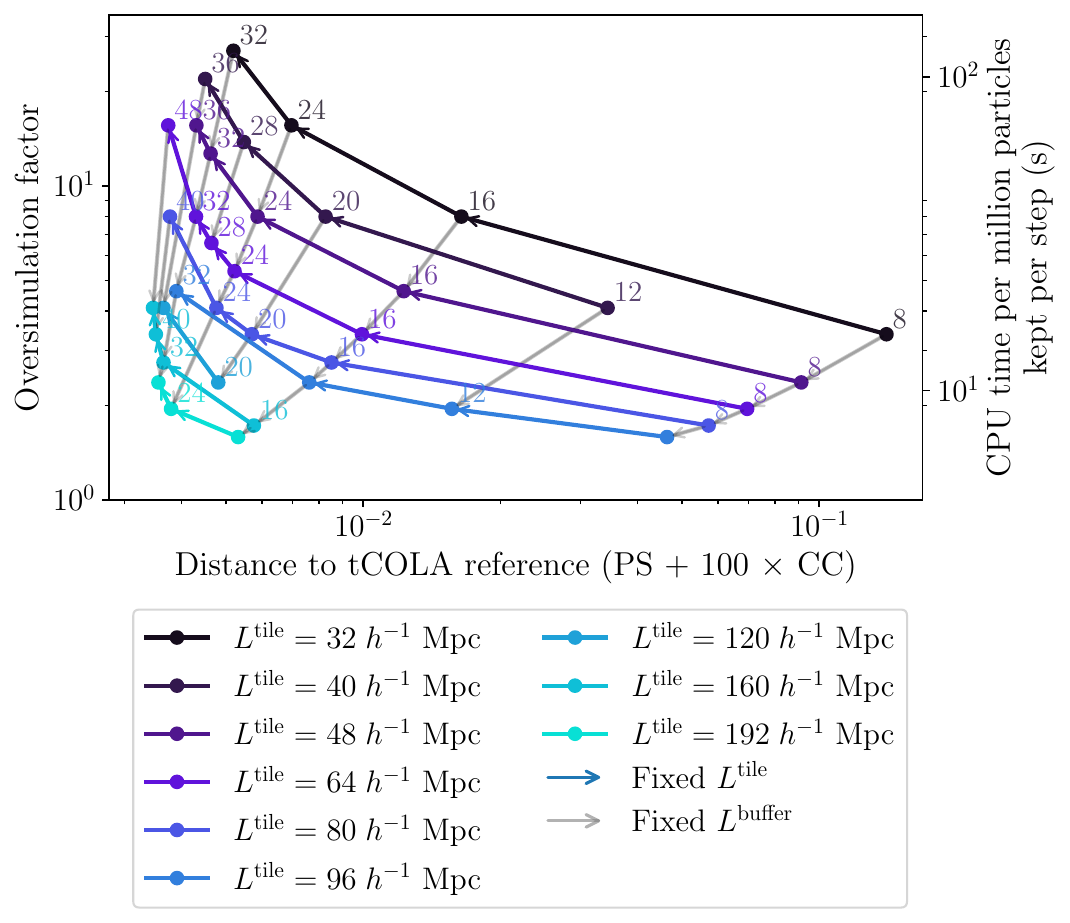}
        \caption{Pareto front of the sCOLA configurations (tile and buffer sizes), comparing the oversimulation factor (or equivalently the CPU sec / M part kept / step) to the distance to tCOLA, as defined in equation~\eqref{eq:distance_tCOLA}. Coloured lines show varying buffer size at fixed tile size, grey lines show varying tile size at fixed buffer size. The annotated coloured labels correspond to the buffer sizes (in $h^{-1}\,\mathrm{Mpc}$). At a given accuracy, the largest tiles minimise the oversimulation factor. 
        }
        \label{fig:pareto_oversimulation}
    \end{figure}

\begin{figure}
        \vspace{0.5cm}
        \centering
        \includegraphics[width=1\linewidth]{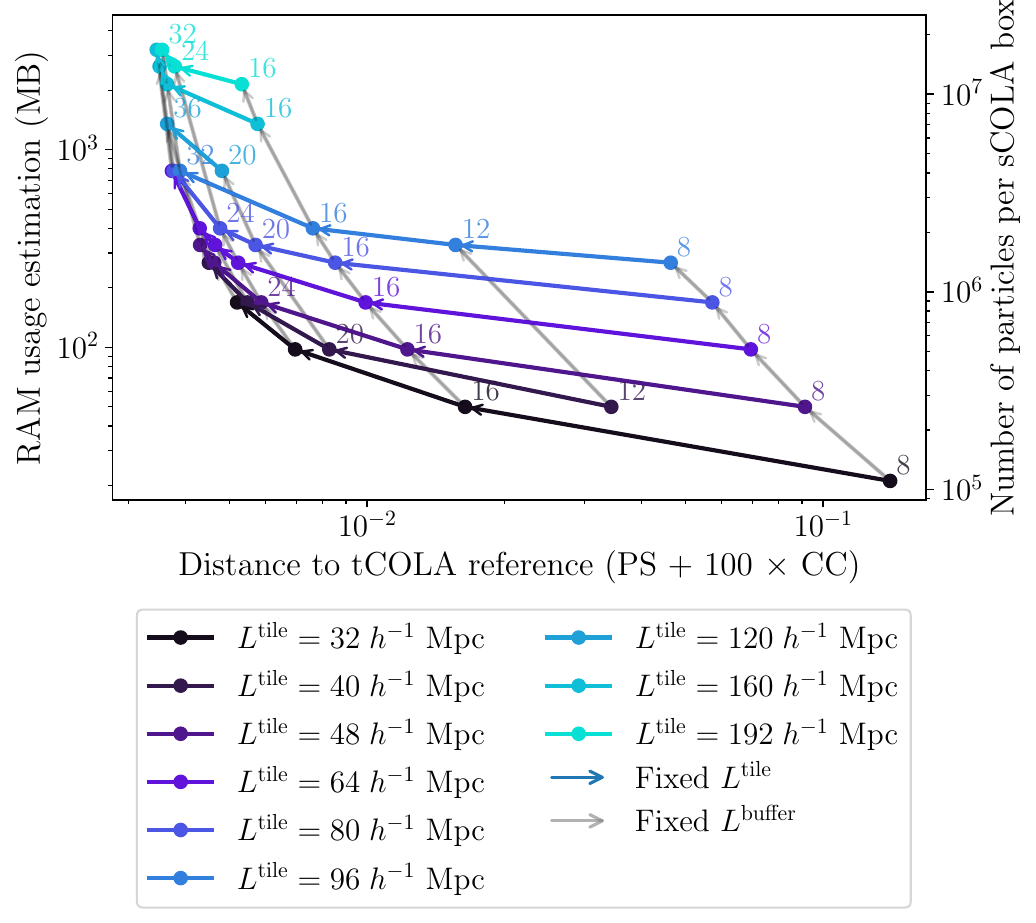}
        \caption{Pareto front of the sCOLA configurations (tile and buffer sizes), comparing the RAM usage (or equivalently the number of particles in the sCOLA box) to the distance to tCOLA (as defined in equation~\eqref{eq:distance_tCOLA}). Coloured lines show varying buffer size at fixed tile size, grey lines show varying tile size at fixed buffer size. The annotated coloured labels correspond to the buffer sizes (in $h^{-1}\,\mathrm{Mpc}$). At a given accuracy, the smallest tiles minimise the RAM usage.}
        \label{fig:pareto_ram}
    \end{figure}

    For $L^\mathrm{tile}\in [30.0,200.0]\,h^{-1}\,\mathrm{Mpc}$, $L^\mathrm{buffer}\in [8.0,48.0]\,h^{-1}\,\mathrm{Mpc}$, and the \textit{Micro} volume defined above ($d_\mathrm{p}=1\,h^{-1}\,\mathrm{Mpc}$, $d_\mathrm{pm}=0.5\,h^{-1}\,\mathrm{Mpc}$), 72 sets of parameters satisfy these constraints. Running all of them is unnecessary: the quantities we are after --- the trend of the distance to tCOLA with $L^\mathrm{tile}$ and $L^\mathrm{buffer}$, and the Pareto fronts these trace out --- are smooth in both parameters, so a coarser sampling resolves them just as well. We therefore ran 35 configurations, thinning the $(L^\mathrm{tile},L^\mathrm{buffer})$ grid while keeping the full extent of both ranges. Figures~\ref{fig:sCOLA_tile_size} and \ref{fig:sCOLA_buffer_size} show the power spectra and cross-correlations of a subset of these configurations compared to tCOLA; Figs.~\ref{fig:pareto_oversimulation} and \ref{fig:pareto_ram} show, for all of them, the Pareto fronts of the distance to tCOLA against the oversimulation factor and the RAM usage.

    We compare the distance $D_\mathrm{ref}$ to tCOLA, defined in equation~\eqref{eq:distance_tCOLA}, to two cost measures:
the oversimulation factor of equation~\eqref{eq:oversimulation}, which is equivalent to the CPU seconds per million particles kept per step (CPU sec / M part kept / step); and the RAM required, which is equivalent to the number of particles in the sCOLA box, since both the particle arrays and the PM grids scale with the number of particles. The first equivalence holds if the cost per million particles evolved per step, counting those of the buffer, is the same for every configuration. We did not time the \textit{Micro} configurations individually, so we take this as an assumption; its scale is set by the only sCOLA timing at our disposal, the ${\sim}4.6$ CPU sec / M part / step measured on the \textit{Main} boxes (Section~\ref{ssec:Computational cost}).
Plotting each cost measure against the distance metric yields the Pareto fronts of Figs.~\ref{fig:pareto_oversimulation} and \ref{fig:pareto_ram}. Minimising the oversimulation ratio (CPU sec / M part kept / step) and minimising the RAM usage at a given accuracy are conflicting objectives: the Pareto optimum for the former is the largest tiles, whereas for the latter it is the smallest tiles. Increasing the buffer size moves along the Pareto front. The effect on the Pareto fronts of the particle resolution $d_\mathrm{p}$, fixed here to $1.0\,h^{-1}\,\mathrm{Mpc}$ (or equivalently of the mass resolution), is beyond the scope of this article.

    The Pareto fronts of Figs.~\ref{fig:pareto_oversimulation} and \ref{fig:pareto_ram} can be used to choose the most efficient configuration as follows. First, determine the RAM available per sCOLA job (with 4 CPUs per job, the RAM associated with these 4 CPUs, for instance 2~GB). Then, set a target accuracy in terms of the distance metric (for instance, 1\% accuracy corresponds approximately to $D_{\mathrm{ref}}=10^{-2}$). These two constraints delimit an allowed region in the bottom-left corner of the RAM Pareto plot (Fig.~\ref{fig:pareto_ram}): at the mass resolution of this work, 2~GB allow at most $230^3$ particles per sCOLA box. Finally, among the configurations in this region, choose the one with the smallest oversimulation factor in Fig.~\ref{fig:pareto_oversimulation} (the lowest point). Here this is $L^\mathrm{tile}=192\,h^{-1}\,\mathrm{Mpc}$, $L^\mathrm{buffer}=16\,h^{-1}\,\mathrm{Mpc}$, with an oversimulation factor of 1.59, the configuration discussed in Section~\ref{sec:Discussion}. Both Pareto fronts also show where the returns from enlarging the buffer stop: at every tile size, the distance to tCOLA improves sharply up to $L^\mathrm{buffer}\simeq16\,h^{-1}\,\mathrm{Mpc}$ and only marginally beyond, while the oversimulation factor keeps growing. With the LPT buffer and the second-order boundary conditions of Section~\ref{ssec:Improvements of sCOLA} enabled, buffers of $L^\mathrm{buffer} \gtrsim 16\,h^{-1}\,\mathrm{Mpc}$ are therefore sufficient. In this work, we focused on the configuration $L^\mathrm{tile}=64\,h^{-1}\,\mathrm{Mpc}$, $L^\mathrm{buffer}=16\,h^{-1}\,\mathrm{Mpc}$ to demonstrate that such small tiles and buffers reach 1\% accuracy on the power spectrum; larger tiles would however have been more efficient in terms of computing resources.

\section{Equilateral bispectra}\label{app:bispectra}

    \begin{figure*}
        \centering
        \includegraphics[width=\textwidth]{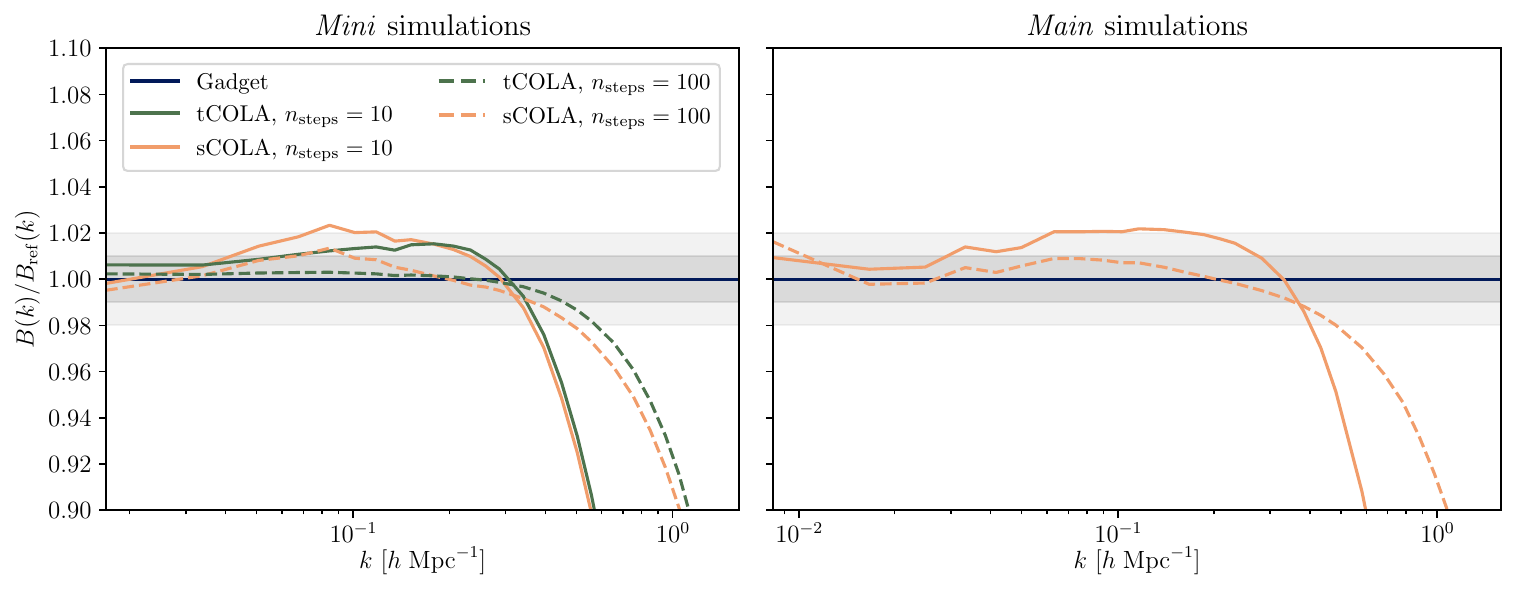}
        \caption{Comparison of the matter density equilateral bispectra between \texttt{Gadget4}, sCOLA and tCOLA on the \textit{Mini} and \textit{Main} lightcone simulations, for 10 and 100 time steps ($n_\mathrm{steps}$ variable). We plot the ratio to the reference \texttt{Gadget4} bispectrum. The grey bands are 1\% and 2\% around \texttt{Gadget4}. sCOLA agrees with \texttt{Gadget4} at the $\sim 2\%$ level up to $k\simeq 0.4\,h\,\mathrm{Mpc}^{-1}$.}
        \label{fig:bispectra}
    \end{figure*}

    We used a fork of the \texttt{DensityFieldTools} code \citep{DensityFieldTools} to compute the equilateral bispectra of the dark matter density fields of the \textit{Mini} and \textit{Main} simulations, using the following estimator:
    \begin{align}
        \delta_{\rm shell}^\mathrm{bin}(\bm x)&\equiv\sum_{|\bm k|\in \mathrm{bin}}\delta(\bm k)\erm^{\irm\bm k\cdot \bm x}\\
        1_{\rm shell}^\mathrm{bin}(\bm x)&\equiv\sum_{|\bm k|\in \mathrm{bin}}1\cdot \erm^{\irm\bm k\cdot \bm x}\\
        N_\triangle^\mathrm{bin}&=N^6\sum_{\bm x}1_{\rm shell}^\mathrm{bin}(\bm x)^3 \\
        B^\mathrm{bin}&=\frac{1}{N_\triangle^\mathrm{bin}}\frac{V^2}{N^3}\sum_{\bm x}\delta_{\rm shell}^\mathrm{bin}(\bm x)^3
    \end{align}
    with $\delta_{\rm shell}^\mathrm{bin}$ the overdensity field restricted to one $k$ bin, computed with a fast Fourier transform and its inverse; $N_\triangle^\mathrm{bin}$ the number of equilateral triangles in the considered bin; $N$ the number of grid cells per dimension used for the Fourier transforms; $V$ the volume of the simulation box; and $B^\mathrm{bin}$ the estimated equilateral bispectrum of this bin.

    Figure~\ref{fig:bispectra} compares the equilateral bispectra of sCOLA, tCOLA and \texttt{Gadget4}. sCOLA agrees with \texttt{Gadget4} at the $\sim 2\%$ level up to $k\simeq 0.4\,h\,\mathrm{Mpc}^{-1}$. The approximations have effects comparable to those observed on the power spectrum: the sCOLA tiling introduces a slight bump around the fundamental mode of the sCOLA box, $k_\mathrm{sCOLA}$; the reduced number of time steps leads to an overall increase of the bispectrum with a divergence at small scales; and the PM force model results in a divergence around $k\sim1\,h\,\mathrm{Mpc}^{-1}$.

\section{Velocity power spectra}\label{app:velocity_ps}

    \begin{figure*}
        \centering
        \includegraphics[width=\textwidth]{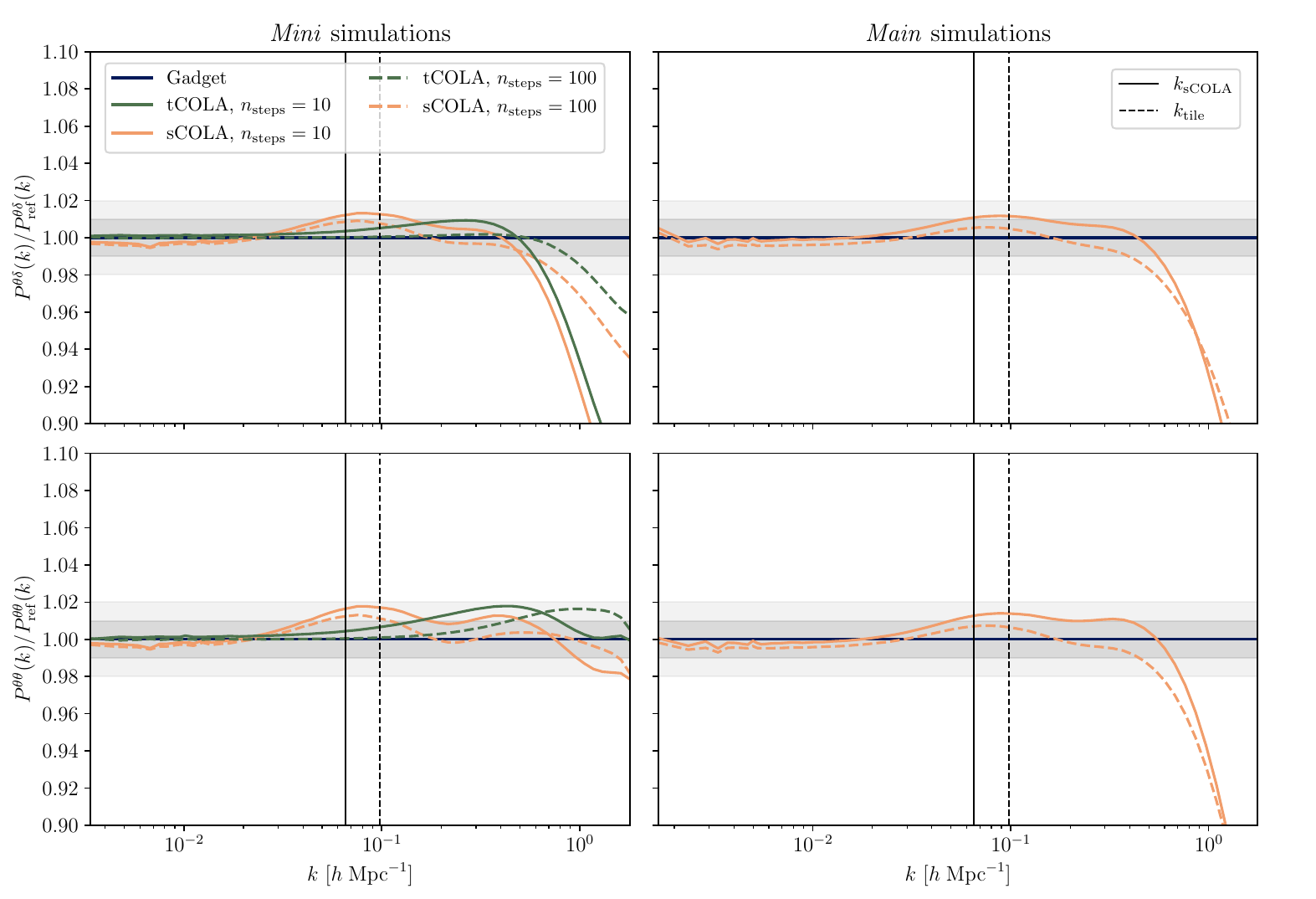}
        \caption{Velocity statistics of the \textit{Mini} (left column) and \textit{Main} (right column) lightcone simulations, for \texttt{Gadget4}, tCOLA and sCOLA. The top row shows the cross-spectrum $P^{\theta\delta}$ between the velocity divergence $\theta$ and the matter density $\delta$, the bottom row the divergence auto-spectrum $P^{\theta\theta}$; both are plotted as ratios to the \texttt{Gadget4} reference, and the grey bands mark $1\%$ and $2\%$ around it. Solid lines are runs with $n_\mathrm{steps}=10$ time steps and dashed lines runs with $n_\mathrm{steps}=100$; the monolithic tCOLA reference is available for the \textit{Mini} volume only (Section~\ref{ssec:Monolithic tCOLA}). sCOLA agrees with \texttt{Gadget4} at the $1.5\%$ level up to $k\simeq0.6\,h\,\mathrm{Mpc}^{-1}$.}
        \label{fig:ps_velocity}
    \end{figure*}

    We deposit the lightcone velocity field on a grid with the Simplex-in-Cell (SiC) projection \citep{HahnAnguloAbel2015,Leclercq2017DMSHEET} already used in Section~\ref{ssec:Density and velocity fields}, on the same grid as the density field, and compute its divergence $\theta$ in Fourier space with fast Fourier transforms.
Figure~\ref{fig:ps_velocity} compares the velocity divergence power spectra of sCOLA, tCOLA and \texttt{Gadget4}, together with their cross-correlations with the overdensity field $\delta$. The density--divergence cross-correlations $P^{\theta\delta}$ and the divergence power spectrum $P^{\theta\theta}$ of sCOLA agree with \texttt{Gadget4} at the 1.5\% level up to
$k\simeq 0.6\,h\,\mathrm{Mpc}^{-1}$.
    For both $P^{\theta\delta}$ and $P^{\theta\theta}$, the effects of the approximations are the same as on the matter power spectrum and bispectrum.

\section{Volume and cost of a \textit{Euclid} north cap simulation}\label{app:northcap}

    This appendix derives the two numbers quoted in Section~\ref{ssec:Lightcone simulations and applications}: the volume of the \textit{Euclid} Galactic north cap relative to its bounding cube, and the cost of simulating that region with sCOLA.

    We model the north cap as a truncated spherical cone sector. With the cosmological parameters of Section~\ref{ssec:General setup}, the redshift range $0.84\leq z\leq 1.88$ translates into the comoving range $R_\mathrm{min} \leq R \leq R_\mathrm{max}$, with $R_\mathrm{min}=2015\,h^{-1}\,\mathrm{Mpc}$ and $R_\mathrm{max}=3470\,h^{-1}\,\mathrm{Mpc}$, and the Galactic latitude cut $b > b_\mathrm{max} = 23^\circ$ into the opening angle $\theta_\mathrm{max} = 90^\circ - b_\mathrm{max} = 67^\circ$. Integrating $2\pi\,r^2\sin\theta$ over the sector gives its volume,
    \begin{equation}
        \begin{aligned}
            V &= \frac{2\pi}{3}\left(R_\mathrm{max}^3-R_\mathrm{min}^3\right)\left(1-\cos\theta_\mathrm{max}\right) \\
              &= 42.9\,(h^{-1}\,\mathrm{Gpc})^3 .
        \end{aligned}
        \label{eq:northcap_volume}
    \end{equation}
    Since the sector has an axis of symmetry, it fits most tightly in a cube whose main diagonal carries that axis. Let $\alpha \equiv \arccos(1/\sqrt{3}) \simeq 54.7^\circ$ be the angle between that diagonal and any edge of the cube. Since $\theta_\mathrm{max} > \alpha$, the direction of each edge falls inside the sector, which therefore reaches the distance $R_\mathrm{max}$ along it; in the opposite direction it extends only to $-R_\mathrm{max}\cos(\theta_\mathrm{max}+\alpha)$, on its rim. The side of the bounding cube is the sum of the two,
    \begin{equation}
        L_\mathrm{cube}= R_\mathrm{max}\left[1-\cos\left(\theta_\mathrm{max}+\alpha\right)\right] .
        \label{eq:northcap_cube}
    \end{equation}
    We find $L_\mathrm{cube} = 5295\,h^{-1}\,\mathrm{Mpc}$, so that the cap occupies a fraction $s_\mathrm{survey} = V/L_\mathrm{cube}^3 = 29\%$ of its bounding cube.

    The cost of simulating the cap with sCOLA follows. Only the tiles overlapping the footprint need to be evolved, which saves the factor $s_\mathrm{survey}$, and over this redshift range the lightcone reduces the average number of time steps by a further factor $s_\mathrm{lightcone} \simeq 45\%$, estimated from Fig.~\ref{fig:fraction_tiles}. Writing $N_\mathrm{cube}=(L_\mathrm{cube}/d_\mathrm{p})^3$ for the number of particles in the bounding cube at the particle resolution $d_\mathrm{p}=1\,h^{-1}\,\mathrm{Mpc}$, and $c$ for the cost per particle and per time step, the cost of one north cap simulation is
    \begin{equation}
        C = s_\mathrm{survey}\,s_\mathrm{lightcone}\,r\,N_\mathrm{cube}\,n_\mathrm{steps}\,c .
        \label{eq:northcap_cost}
    \end{equation}
    With $n_\mathrm{steps}=10$, the oversimulation factor $r\simeq1.6$ of the $L^\mathrm{tile}=192\,h^{-1}\,\mathrm{Mpc}$, $L^\mathrm{buffer}=16\,h^{-1}\,\mathrm{Mpc}$ configuration (equation~\eqref{eq:oversimulation}), and $c=1.28\times10^{-9}$~CPUh, which is the $4.6$ CPU seconds per million particles per time step measured in Section~\ref{ssec:Computational cost}, this gives $C \simeq 400$~CPUh.

\section*{Statement of contribution}

MA implemented the lightcone construction, the LPT buffer and the second-order boundary conditions in the sCOLA module of \texttt{Simbelmynë}, ran the simulations, performed the analysis, produced the figures, and wrote the first draft of the paper.
FL conceived and designed the study, wrote the \texttt{Simbelmynë} code and the original sCOLA algorithm, did the analytic calculation of the second-order gravitational potential approximation term $\Phi_2$, contributed to the analysis and interpretation of the results, edited the manuscript, supervised MA, and secured funding.
GL contributed to the design of the algorithm, in particular the LPT buffer, co-supervised MA, contributed to the interpretation of the results, contributed to writing the paper, and secured funding.
All authors read and approved the final manuscript.

\section*{Declaration of generative AI use}

Generative AI assistants were used to help develop and review parts of the \texttt{Simbelmynë} code, and to improve the clarity, style, and grammar of the manuscript. All AI-assisted code was reviewed by the authors and subjected to the same testing and validation procedures as manually written code. The authors take responsibility for the scientific content of the paper.

\section*{Acknowledgements}

We thank Oliver Hahn, Andrew Jaffe, Cornelius Rampf and Benjamin Wandelt for useful discussions, and Ludvig Doeser for comments on the draft. This work was done within the Aquila Consortium (\url{https://www.aquila-consortium.org/}).

This work has made use of the Infinity Cluster hosted by the Institut d'Astrophysique de Paris; we appreciate the efforts of Stéphane Rouberol for running the cluster smoothly. Infinity was supported by the \textit{Domaine d'Intérêt Majeur} (DIM) ORIGINES of the Région Île-de-France. This work was granted access to the resources of the Très Grand Centre de calcul du CEA (TGCC) under the allocation A0190416910 made by GENCI.
MA is funded by an AMX doctoral fellowship from the École polytechnique. MA, FL and GL acknowledge financial support from the Agence Nationale de la Recherche (ANR) through the grant INFOCW, under reference ANR-23-CE46-0006-01. FL and GL acknowledge financial support from the Centre National d'Études Spatiales (CNES), and from the Simons Collaboration on \href{https://www.learning-the-universe.org/}{``Learning the Universe''}.

This work made use of \texttt{Gadget4} \citep[\href{https://wwwmpa.mpa-garching.mpg.de/gadget4/}{wwwmpa.\allowbreak mpa-garching.\allowbreak mpg.\allowbreak de/\allowbreak gadget4},][]{Gadget4}, \texttt{MUSIC2-monofonIC} \citep[\href{https://github.com/cosmo-sims/monofonIC}{github.\allowbreak com/\allowbreak cosmo-sims/\allowbreak monofonIC},][]{monofonIC}, \texttt{CLASS} \citep[\href{http://class-code.net}{class-code.\allowbreak net},][]{CLASS} and \texttt{Simbelmynë} \citep[\href{https://git.aquila-consortium.org/Aquila-Consortium/simbelmyne}{git.\allowbreak aquila-consortium.\allowbreak org/\allowbreak Aquila-Consortium/\allowbreak simbelmyne},][]{Leclercq2015ST}, as well as \texttt{NumPy} \citep{NumPy}, \texttt{SciPy} \citep{SciPy} and \texttt{Matplotlib} \citep{Matplotlib}. The figures use the scientific colour maps of \citet{Crameri_colourmaps} and the \texttt{CMasher} package \citep{CMasher}.

\section*{References}
\bibliography{refs}

\end{document}